\documentclass[a4paper,usenatbib]{mnras}
\usepackage{graphicx}
\usepackage[fleqn]{amsmath}
\usepackage[T1]{fontenc}
\usepackage{aecompl}
\usepackage{times}
\usepackage{ae,aecompl}
\usepackage{amssymb,amsfonts,hyperref,color}

\newcommand\msunyr{\rm M_{\odot}\,yr^{-1}}
\newcommand\be{\begin{equation}}
\newcommand\en{\end{equation}}

\newcommand\etal{{\rm et al}.\ }

\title[Circumplanetary Disc Detection]{On the Radio Detectability of Circumplanetary Discs}

\author[Zhu, Andrews \& Isella]{Zhaohuan Zhu$^{1}$\thanks{E-mail: zhaohuan.zhu@unlv.edu}
Sean M. Andrews$^{2}$ and Andrea Isella$^{3}$\\
$^{1}$Department of Physics and Astronomy, University of Nevada, Las Vegas, 
                4505 S. Maryland Pkwy, Las Vegas, NV, 89154, USA\\
$^{2}$Harvard-Smithsonian Center for Astrophysics, 60 Garden Street, Cambridge, MA 02138, USA\\
$^{3}$Department of Physics and Astronomy, Rice University, 6100 Main Street, Houston, TX 77005, USA}
\date{In original form \today}

\begin{document}
\label{firstpage}
\pagerange{\pageref{firstpage}--\pageref{lastpage}} \pubyear{2017}
\maketitle

\begin{abstract}
Discs around young planets, so-called circumplanetary discs (CPDs), are essential for planet growth, satellite formation, and planet detection. 
We study the millimetre and centimetre emission from accreting CPDs by using the simple $\alpha$ disc model. We find that it is easier to
detect CPDs at shorter radio wavelengths (e.g. $\lambda\lesssim$ 1 mm). For example, if the system is 140 pc away from us, 
deep observations (e.g. 5 hours) at ALMA Band 7 (0.87 mm) are sensitive to as small as 0.03 lunar mass of dust in CPDs. If the CPD is around a Jupiter mass 
planet 20 AU away from the host star and has $\alpha\lesssim 0.001$, 
ALMA can detect this disc when it accretes faster than $10^{-10} M_{\odot}/yr$. ALMA can also detect the ``minimum mass sub-nebulae'' disc
if such a disc exists around a young planet in YSOs.  However, to distinguish the embedded compact CPD from the circumstellar disc material, 
we should observe circumstellar discs with large gaps/cavities using the highest resolution possible. We also calculate the CPD fluxes at VLA  bands, and
discuss the possibility of detecting radio emission from jets/winds launched in CPDs. 
Finally we argue that, if the radial drift of dust particles is considered, the drifting timescale for millimetre dust in CPDs can be extremely short.
It only takes 10$^2$-10$^{3}$ years for CPDs to lose millimetre dust. Thus, for CPDs to be detectable at radio wavelengths, mm-sized dust in 
CPDs needs to  be replenished continuously, or the disc has a significant fraction
of micron-sized dust or a high gas surface density so that the particle drifting timescale is long, or  the radial drift is prevented by other means  (e.g. pressure traps). 
\end{abstract}

\begin{keywords}
radiation mechanisms: thermal --- planets and satellites: detection  --- planet-disc interactions  --- protoplanetary discs --- brown dwarfs  --- 
radio continuum: planetary systems --- submillimetre: planetary systems
\end{keywords}

\section{Introduction}
{During their formation, giant planets are expected to be surrounded by dense discs made of gas and dust called
circumplanetary discs (CPDs).} 
The orbital properties of regular satellites around giant planets in our solar system support the existence of CPDs in the past.
All medium and large satellites of giant planets in our solar systems, except the captured satellite Triton, have prograde orbits that are nearly coplanar with the planet's equator. 
Furthermore, the gradient in the amount of ices in the Galilean satellites is consistent with satellite formation in a warm disc.  

Theoretical studies suggest that, after the runaway phase of giant planet formation, the planet's atmosphere detaches from the planet's Hill sphere and shrinks significantly \citep{PapaloizouNelson2005}. 
Material that falls into the planet's Hill sphere will spin around the planet, forming CPDs to conserve the angular momentum \citep{Lubow1999, AyliffeBate2009, Szulagyi2016}. The planet can continue
to grow with material being accreted through the CPDs.

{Studying CPDs is crucial for understanding
satellite formation, and even habitability of life on satellites (e.g. Europa and Enceladus, \citealt{Waite2017})}. The density and temperature
structure of CPDs affect the satellite size and composition. If the CPD is too hot for water to condense, the satellites will be made mostly of dust grains
with little subsurface water ocean \citep{CanupWard2002}. Forming large satellites and multiple satellites in resonance (e.g. Galilean satellites) can lead to geological activity, providing
energy sources for early life \citep{Peale1999}. 
 
Despite their importance, finding CPDs in other young stellar objects (YSOs) is challenging since these discs are a lot fainter than the central star.
On the other hand, when CPDs are actively accreting, they can be as bright as a late M-type/early L-type brown dwarf  \citep{QuillenTrilling1998,Zhu2015}.
Unlike the SEDs of a brown dwarf, the SEDs of accreting CPDs are a lot redder, peaking at mid-infrared. 
Thus, to detect CPDs in protostellar systems,
we need to image the young stellar system at mid-infrared, adopting high contrast imaging techniques capable of blocking most of the stellar light.
With such observational strategy, a few red sources within circumstellar discs have been discovered 
(LkCa 15b: \citealt{KrausIreland2012}, HD100546b: \citealt{Quanz2013, Currie2015}, but see \citealt{Rameau2017}, HD169142b: \citealt{Biller2014, Reggiani2014}) and their photometries at 
infrared bands are consistent with accreting CPDs \citep{Zhu2015, Zhu2016}. Besides the photometry,
another observational signature of accretion discs is 
$H_{\alpha}$ emission lines \citep{CalvetGullbring1998, Muzerolle1998, Muzerolle2001, Zhou2014}.
 So far, LkCa 15b is the most promising candidate for the accreting CPDs 
since both the accretion indicators, H$_{\alpha}$ line and near-IR thermal
emission, have been detected \citep{Sallum2015} .
 
Besides optical and infrared observations, radio observations can potentially detect the CPDs too. Using the parameterized models, 
 \cite{Isella2014} suggests that ALMA can detect CPDs with mass down to $5\times 10^{-4}$ $M_{J}$ and size down to 0.2 AU. 
With numerical simulations, \cite{Zhu2016} suggests that the central star can induce spiral shocks in CPDs and the
dissipation of the shocks can lead to CPD accretion.
Applying this accretion disc model to the CPD candidates mentioned above (HD100546b, HD169142b, LkCa 15b), all these CPD candidates 
should be detectable by ALMA \citep{Zhu2016} .
In this paper, we extend  \cite{Isella2014} with detailed disc vertical structure calculations to show whether
CPDs can be detected by ALMA and future generation radio telescopes such as the Next Generation VLA (ngVLA, \citealt{SelinaMurphy2017}).
In \S 2, we introduce the analytical disc model. The results are presented in \S 3. 
Finally, after short discussions in \S 4, the paper is
summarized in \S 5.

\section{Method}
To calculate the radio emission from CPDs, we first need to construct a disc model. 
Following the $\alpha$ disc theory, and similarly to circumstellar discs \citep{DAlessio1998}, the structure of a CPD is determined by 
the disc accretion rate
($\dot{M_{p}}$), the viscosity parameter ($\alpha$), and the disc inner and outer radius ($R_{in}$, $R_{out}$), as long as 
the planet mass ($M_{p}$), temperature ($T_{p}$), and radius (R$_{p}$) are given.
Unfortunately, we don't have much, if any, observational constraints on these parameters. Thus, we narrow 
down the parameter space 
using either order-of-magnitude estimates or theoretical results from numerical calculations.

To form a Jupiter mass planet within the lifetime of
the gaseous circumstellar discs ($\sim10^6$ yrs), the average CPD accretion rate ($\dot{M_{p}}$) is
$10^{-6}$ M$_{J}$/yr. In this paper,
we vary $M_{p}\dot{M_{p}}$ from $10^{-5}$ M$_{J}^2$/yr to $10^{-8}$ M$_{J}^2$/yr.  We choose $M_{p}\dot{M_{p}}$
as a fundamental parameter since
$M_{p}\dot{M_{p}}$ determines the disc effective temperature for a viscous heating dominated disc (Equation \ref{eq:Fv}). However, for the
radio emission calculation, 
$M_{p}$ and $\dot{M_{p}}$ become degenerate when the disc is optically thin or irradiation dominated (\S 3.1, 3.2, 4.1).
Thus, we normally also specify $M_{p}$ whenever $M_{p}\dot{M_{p}}$ is given. 

Another important parameter for determining the disc structure is the viscosity coefficient: $\alpha$ parameter.
The first-principle hydrodynamical 
simulations by \cite{Zhu2016}  suggest that spiral shocks induced by the star
can lead to efficient angular momentum transport in CPDs. In this accretion scenario, the value of $\alpha$ parameter depends
on the disc accretion rate due to the thermal feedback. For a disc accreting at a higher rate, the disc is hotter, leading to more-opened
 spiral shocks so that the shock-driven accretion is more efficient with a higher $\alpha$.
Over a wide range of accretion rates ($\dot{M}_{p}$ from $10^{-12}M_{\odot}/yr$ to  $10^{-8}M_{\odot}/yr$),
 $\alpha$ can vary from $0.0001$ to $0.1$.

The disc inner radius is assumed to be the Jupiter radius ($R_{J}$), and
the outer radius of the CPD ($R_{out}$) is assumed to be
1/3 of the Hill radius of the planet \citep{QuillenTrilling1998,MartinLubow2011}, or
$R_{out}=1/3\times r_{p}(M_{p}/3M_{*})^{1/3}$.
In this paper, we use $R$ to represent the radius in the circumplanetary disc (the distance to the planet), and 
$r$ to represent the radius in the circumstellar disc (the distance to the central star).
For a Jupiter mass 
planet at $r_{p}$=5, 20, and 80 AU from the solar mass star, $R_{out}$ is 0.12, 0.46, and 1.85 AU. 

With these disc parameters ($\dot{M_{p}}$, $\alpha$, $R_{in}$, and $R_{out}$), we can calculate the disc
structure by solving coupled radiative transfer and hydrostatic equilibrium equations as in D'Alessio \etal (1996).
However, such detailed CPD models are not necessary at the current stage where these parameters are still highly uncertain.
Thus  in this paper, we make approximations to simplify the calculations.

Given the $M_{p}\dot{M_{p}}$, the viscous theory suggests that the
disc effective temperature follows
\begin{equation}
T_{eff}^4 = {3 G M_{p}\dot{M_{p}} \over 8\pi\sigma R^{3}}
\left[1-\left(\frac{R_{in}}{R}\right)^{1/2}\right]\,,
\label{eq:Fv}
\end{equation}
where $R_{in}$ is the disc inner radius and we assume $R_{in}=0.1$R$_{\odot}\sim$ R$_{J}$.
To derive the disc midplane temperature, we assume that the disc vertical temperature along the height (z)  
follows \cite{Hubeny1990}, 
\begin{equation}
T(z)^{4}=\frac{3}{8}T_{eff}^4 \tau_{R}+T_{ext}^4\,,\label{eq:Tz}
\end{equation}
where $\tau_{R}$ is the Rosseland mean optical depth at $z$,
 and $T_{ext}$ is the background temperature due to the external irradiation.
Here we include both the heating from ISM and the irradiation from the planet as
$T_{ext}^4=T_{ISM}^4+T_{irr}^4$. We choose $T_{ISM}=10$ K and assume that 
$T_{irr}$ could be due to the radiation from either the central bright planet or the disc boundary layer  \footnote{We have ignored the irradiation from the protoplanetary disc and the central star. 
Such irradiation can be important if the planet is close to the central star (e.g. 5 AU, see discussion in \S 4.1). On the other hand,
as shown in \S 3.2 and \S 4.1, irradiation makes little difference on the radio emission.}. 
The planet or the boundary layer will irradiate the disc at a large incident angle. 
The incident angle is $tan^{-1}(H/R)$ to the
first order \citep{KenyonHartmann1987}. With $H/R\sim 0.1$ \citep{Zhu2016}, we assume that the disc receives 1/10th of 
the irradiated energy in the direction perpendicular to the disc surface. Thus,
\begin{equation}
T_{irr}=\left(\frac{L_{irr}}{\sigma 40 \pi R^2 }\right)^{1/4}\,.\label{eq:tirr}
\end{equation}
When we consider the irradiation from a bright planet, we assign $L_{irr}$  based on the ``hot start'' planet model (e.g. \citealt{Marley2007, SpiegelBurrows2012}).
When we consider the irradiation from the boundary layer, we adopt
$L_{irr} =GM_{p}\dot{M_{p}}/(2 R_{p})$ since half of the accretion energy is released 
at the boundary layer.

The disc midplane temperature is then
\begin{equation}
T_{c}^4 = {9 G M_{p}\dot{M_{p}}\Sigma\kappa_{R} \over 128\pi\sigma R^{3}}
\left[1-\left(\frac{R_{in}}{R}\right)^{1/2}\right]+T_{ext}^4\,,\label{eq:Tc}
\end{equation}
where $\kappa_{R}$ is the Rosseland mean opacity and we choose the nominal value
of $\kappa_{R}=10$ cm$^2$/g. 

Using the viscous disc theory, we also have
\begin{equation}
\nu\Sigma={\dot{M} \over 3\pi}
\left[1-\left(\frac{R_{in}}{R}\right)^{1/2}\right]\,,
\label{eq:nusigma}
\end{equation}
where $\nu=\alpha c_{s, iso}^2/\Omega$ in the $\alpha$ disc model.

With Equations \ref{eq:Fv}-\ref{eq:nusigma}, we can solve $\Sigma$ numerically. On the other hand, 
we can also approximate the solution with analytical formulae in the following way. First, if we ignore $T_{ext}$, we can solve
$\Sigma$ analytically as  
\begin{equation}
\Sigma=\frac{2^{7/5}}{3^{6/5}}\left(\frac{\sigma GM_{p}\dot{M_{p}}^{3}}{\alpha^4\pi^3 \kappa_{R} R^3}\right)^{1/5}\left(\frac{\mu}{\Re}\right)^{4/5}\left[1-\left(\frac{R_{in}}{R}\right)^{\frac{1}{2}}\right]^{3/5}\,,\label{eq:Sigma}
\end{equation}
where $\mu$ and $\Re$ are the mean molecular weight and the gas constant. Considering the disc is mostly cold, we choose $\mu=2.4$.
On the other hand, if we ignore the viscous heating and assume $T_{c}$ is dominated by $T_{ext}$, we can solve
$\Sigma$ as
\begin{equation}
\Sigma=\frac{\dot{M_{P}}\mu\Omega}{3\pi\alpha \Re T_{ext}}\,.\label{eq:Sigma2}
\end{equation}
In reality, both viscous heating and external irradiation contribute to $T_{c}$, and
a larger $T_{c}$ will result in a smaller $\Sigma$ (Equation \ref{eq:nusigma}).
Thus, we choose $\Sigma$ as the minimum
of $\Sigma$ derived from Equations \ref{eq:Sigma} and \ref{eq:Sigma2}. With the $\Sigma$ determined, we can calculate the
midplane temperature using Equation \ref{eq:Tc}.

Knowing the disc surface density and the vertical temperature structure (Equation \ref{eq:Tz}),  
we can finally calculate the disc radio continuum emission. 
First, we calculate the disc's optical depth at the given radio wavelength ($\tau_{mm}=\kappa_{mm}\Sigma/2$).
Although the disc is normally optically thick with the Rosseland mean opacity, it
can be optically thin at  millimetre (mm) and centimetre (cm)  considering the significantly lower dust opacity at  the radio wavelength range. 
We adopt the mass opacity from Andrews \etal (2012), 
$\kappa_{mm}=0.034\times(0.87 mm/\lambda)$ cm$^2$/g which assumes a dust-to-gas mass ratio of 0.01.
When the disc is optically thick at mm, the disc's effective brightness temperature at mm ($T_{b}$)  
\footnote{ In this paper, we define the effective brightness temperature as
the equivalent temperature a black body would have in order to be as bright as the observed brightness, to be distinguished
from the brightness temperature used in the radio literature which refers to the equivalent temperature measured within one observational beam.} will
equal the disc's temperature at $\tau_{mm}(z)=1$.
When the disc is optically thin at mm, we can calculate the
effective brightness temperature by integrating the  product of the temperature, opacity, and density along the disc height. 
Considering most of the disc mass is at the midplane and most radio emission comes from the material at the midplane in the optically thin limit,
the effective brightness temperature is then simply the disc's midplane temperature weighted by the disc optical depth. 

Thus, 
\begin{equation}
T_{b} =
  \begin{cases}
    \left(\frac{3}{8}\frac{\kappa_{R}}{\kappa_{mm}}T_{eff}^4+T_{ext}^4\right)^{1/4}
    \quad \text{if } \tau_{mm}>0.5\\
    \\
    \\ 2T_{c}\tau_{mm}
    \quad \text{if } \tau_{mm}<0.5\\
  \end{cases}\label{eq:Tbr}
\end{equation}
where $\tau_{mm}=0.5\Sigma\kappa_{mm}$.
Then, we can calculate the flux emitted at every disc radius, using $I=2 k T_{b}/\lambda^2$. After integrating the whole disc,
the total CPD emission at mm is derived. We want to caution that, by using the opacity law above and applying the same opacity law throughout the disc, 
we assume that the
dust-to-gas mass ratio is 1/100 and the dust is perfectly coupled to the gas. As shown in \S 4.6, the perfect coupling is only correct for micron sized particles. 
If particles are large enough to decouple from the gas, we could have dramatically different conclusions (\S 4.6). 

we can also calculate the averaged effective brightness temperature 
at mm over the whole CPD surface,
\begin{equation}
\overline{T_{b}}=\frac{\int_{R_{in}}^{R_{out}}T_{b}2\pi RdR}{\pi R_{out}^2}\,.
\end{equation}
This averaged effective brightness temperature can be compared with the brightness temperature limit of the telescope
to estimate if the disc is detectable by ALMA or ngVLA. For example, let's assume 1-hour ALMA band 7 observation has a 0.7 K noise level, and the desired observational beam size is 0.03''.
And we want to observed a CPD  which is 20 AU from the central star and 140 pc away from us. The angular size of the CPD will be 0.006'', which will be smaller than the observational beam by a factor of 5.
We also know that the predicted averaged effective brightness temperature for this CPD is 380 K from Table 1.
Then, the brightness temperature received by one observational beam will be 380/$5^2$=15.2 K, and we can get a S/N=(15.2/0.7)=22 detection with 1-hour integration. 

\begin{table*}
\tiny
\begin{center}
\caption{Viscous Discs with $M_{p}=1 M_{J}$. The left three columns are input parameters, while other columns are derived quantities. \label{tab1}}
\setlength\tabcolsep{3.pt}
\begin{tabular}{ccc|cc|cc|cc|cc|cc|cc|cc|cc|cc}
\hline
\hline
$M_{p}\dot{M_{p}}$&$\alpha$&$R_{out}$& $M_{disc}$ & $L$ & \multicolumn{2}{|c}{F(0.85mm)}  & \multicolumn{2}{|c}{F(1.3mm)} & \multicolumn{2}{|c}{F(3mm)}  & \multicolumn{2}{|c}{F(10mm)} & \multicolumn{2}{|c}{$\overline{T_{b}}(0.85mm)$} & \multicolumn{2}{|c}{$\overline{T_{b}}(1.3mm)$} & \multicolumn{2}{|c}{$\overline{T_{b}}(3mm)$} & \multicolumn{2}{|c}{$\overline{T_{b}}(10mm)$} \\
M$_{J}^2$/yr&&AU&M$_{J}$&$10^{-3}L_{\odot}$ & $\mu$Jy & $\mu$Jy& $\mu$Jy & $\mu$Jy & $\mu$Jy & $\mu$Jy & $\mu$Jy & $\mu$Jy & K & K& K& K & K& K & K & K\\
& & & & & no b. & b. & no b. & b. &  no b. & b. & no b. & b. &  no b. & b. & no b. & b.  & no b. & b. & no b. & b.\\
\hline
1e-5 &      1e-1 &       0.12 &    8.7e-5 &      1.570 &    4.2e1&    4.3e1&    1.3e1&    1.3e1&    1.2&    1.2&    3.3e-2&    3.3e-2 &        219&        225&        159&        163&         76&         78&         24&         24\\
     1e-5 &      1e-2 &       0.12 &    5.5e-4 &      1.570&    7.3e1&    7.4e1&    3.5e1&    3.5e1&    7.3&    7.4&    3.0e-1&    3.0e-1&        380&        384&        422&        425&        475&        477&        213&        214\\
     1e-5 &      1e-3 &       0.12 &    3.5e-3 &      1.570&    7.3e1&    7.4e1&    3.5e1&    3.5e1&    8.0&    8.1&    9.8e-1&    9.8e-1&        380&        384&        422&        425&        520&        522&        703&        704\\
     1e-5 &      1e-4 &       0.12 &    2.2e-2 &      1.570&    7.3e1&    7.4e1&    3.5e1&    3.5e1&    8.0&    8.1&    9.8e-1&    9.8e-1&        380&        384&        422&        425&        520&        522&        703&        704\\
     1e-5 &      1e-1 &       0.46 &    6.1e-4 &      1.570&    1.0e2&    1.2e2&    3.0e1&    3.4e1&    2.5&    2.9&    7.0e-2&    7.9e-2&         33&         38&         23&         26&         10&         12&          3&          4\\
     1e-5 &      1e-2 &       0.46 &    3.8e-3 &      1.570&    4.2e2&    4.4e2&    1.8e2&    1.9e2&    2.1e1&    2.2e1&    6.7e-1&    6.9e-1&        136&        142&        140&        145&         85&         88&         30&         31\\
     1e-5 &      1e-3 &       0.46 &    2.4e-2 &      1.570&    4.2e2&    4.4e2&    2.0e2&    2.0e2&    4.6e1&    4.7e1&    4.5&    4.6 &        136&        142&        151&        155&        186&        189&        204&        205\\
     1e-5 &      1e-4 &       0.46 &    1.5e-1 &      1.570&    4.2e2&    4.4e2&    2.0e2&    2.0e2&    4.6e1&    4.7e1&    5.6&    5.6 &        136&        142&        151&        155&        186&        189&        252&        253\\
     1e-5 &      1e-1 &       1.85 &    4.2e-3 &      1.570&    2.3e2&    2.9e2&    6.4e1&    8.2e1&    5.3&    6.8&    1.5e-1&    1.9e-1&       4.57&       5.90&       3.05&       3.92&       1.35&       1.73&       0.41&       0.52\\
     1e-5 &      1e-2 &       1.85 &    2.7e-2 &      1.570&    1.6e3&    2.0e3&    5.2e2&    6.3e2&    4.9e1&    5.7e1&    1.4&    1.6&         33&         40&         25&         30&         12&         14&          4&          5\\
     1e-5 &      1e-3 &       1.85 &    1.7e-1 &      1.570&    2.4e3&    2.7e3&    1.1e3&    1.2e3&    2.6e2&    2.7e2&    1.2e1&    1.3e1 &         48&         55&         54&         59&         66&         69&         34&         35\\
     1e-5 &      1e-4 &       1.85 &    1.1 &      1.570&    2.4e3&    2.7e3&    1.1e3&    1.2e3&    2.6e2&    2.7e2&    3.2e1&    3.2e1 &         48&         55&         54&         59&         66&         69&         90&         91\\
\hline
     1e-6 &      1e-1 &       0.12 &    2.2e-5 &      0.157&    5.2&    5.6&    1.5&    1.6&    1.2e-1&    1.3e-1&    3.3e-3&    3.5e-3 &         27&         29&         18&         19&          8&          8&          2&          3\\
     1e-6 &      1e-2 &       0.12 &    1.4e-4 &      0.157&    3.5e1&    3.6e1&    1.2e1&    1.2e1&    1.1&    1.1&    3.3e-2&    3.3e-2 &        181&        185&        142&        144&         73&         74&         23&         24\\
     1e-6 &      1e-3 &       0.12 &    8.7e-4 &      0.157&    4.1e1&    4.2e1&    2.0e1&    2.0e1&    4.5&    4.5&    2.7e-1&    2.7e-1 &        213&        216&        237&        239&        293&        293&        195&        195\\
     1e-6 &      1e-4 &       0.12 &    5.5e-3 &      0.157&    4.1e1&    4.2e1&    2.0e1&    2.0e1&    4.5&    4.5&    5.5e-1&    5.5e-1 &        213&        216&        237&        239&        293&        293&        395&        396\\
     1e-6 &      1e-1 &       0.46 &    1.5e-4 &      0.157&    1.1e1&    1.4e1&    3.2&    4.0&    2.6e-1&    3.2e-1&    7.0e-3&    8.7e-3 &       3.65&       4.58&       2.41&       3.01&       1.05&       1.31&       0.31&       0.39\\
     1e-6 &      1e-2 &       0.46 &    9.6e-4 &      0.157&    9.5e1&    1.1e2&    2.8e1&    3.1e1&    2.5&    2.7&    7.0e-2&    7.6e-2 &         31&         34&         22&         24&         10&         11&          3&          3\\
     1e-6 &      1e-3 &       0.46 &    6.1e-3 &      0.157&    2.4e2&    2.5e2&    1.1e2&    1.2e2&    1.8e1&    1.9e1&    6.4e-1&    6.5e-1 &         77&         80&         85&         87&         74&         76&         29&         29\\
     1e-6 &      1e-4 &       0.46 &    3.8e-2 &      0.157&    2.4e2&    2.5e2&    1.1e2&    1.2e2&    2.6e1&    2.6e1&    3.2&    3.2 &         77&         80&         85&         87&        105&        106&        142&        142\\
     1e-6 &      1e-1 &       1.85 &    1.1e-3 &      0.157&    2.7e1&    3.1e1&    7.5&    8.8&    6.1e-1&    7.1e-1&    1.6e-2&    1.9e-2 &       0.54&       0.63&       0.36&       0.42&       0.15&       0.18&       0.05&       0.05\\
     1e-6 &      1e-2 &       1.85 &    6.7e-3 &      0.157&    2.2e2&    2.8e2&    6.5e1&    8.0e1&    5.4&    6.7&    1.5e-1&    1.8e-1 &       4.56&       5.66&       3.07&       3.79&       1.38&       1.69&       0.42&       0.51\\
     1e-6 &      1e-3 &       1.85 &    4.2e-2 &      0.157&    1.3e3&    1.5e3&    4.5e2&    5.2e2&    4.6e1&    5.2e1&    1.4&    1.6 &         26&         30&         22&         25&         12&         13&          4&          4\\
     1e-6 &      1e-4 &       1.85 &    2.7e-1 &      0.157&    1.3e3&    1.5e3&    6.4e2&    7.0e2&    1.5e2&    1.5e2&    1.1e1&    1.1e1 &         27&         31&         30&         33&         37&         39&         30&         31\\
     \hline
     1e-7 &      1e-1 &       0.12 &    5.5e-6 &      0.016&    5.3e-1&    6.4e-1&    1.5e-1&    1.8e-1&    1.2e-2&    1.5e-2&    3.3e-4&    3.9e-4 &       2.77&       3.34&       1.81&       2.18&       0.78&       0.95&       0.24&       0.28\\
     1e-7 &      1e-2 &       0.12 &    3.5e-5 &      0.016&    5.0&    5.3&    1.5&    1.5&    1.2e-1&    1.3e-1&    3.3e-3&    3.4e-3 &         26&         27&         18&         19&          8&          8&          2&          2\\
     1e-7 &      1e-3 &       0.12 &    2.2e-4 &      0.016&    2.3e1&    2.3e1&    9.6&    9.7&    1.0&    1.1&    3.2e-2&    3.3e-2 &        120&        121&        116&        117&         68&         68&         23&         23\\
     1e-7 &      1e-4 &       0.12 &    1.4e-3 &      0.016&    2.3e1&    2.3e1&    1.1e1&    1.1e1&    2.5&    2.5&    2.3e-1&    2.3e-1 &        120&        121&        133&        134&        165&        165&        167&        167\\
     1e-7 &      1e-1 &       0.46 &    3.8e-5 &      0.016&    1.2&    1.5&    3.4e-1&    4.2e-1&    2.7e-2&    3.4e-2&    7.4e-4&    9.2e-4 &       0.39&       0.49&       0.26&       0.32&       0.11&       0.14&       0.03&       0.04\\
     1e-7 &      1e-2 &       0.46 &    2.4e-4 &      0.016&    1.1e1&    1.4e1&    3.2&    3.8&    2.6e-1&    3.2e-1&    7.0e-3&    8.5e-3 &       3.63&       4.41&       2.41&       2.92&       1.06&       1.28&       0.32&       0.38\\
     1e-7 &      1e-3 &       0.46 &    1.5e-3 &      0.016&    8.4e1&    9.1e1&    2.6e1&    2.8e1&    2.4&    2.6&    6.9e-2&    7.4e-2 &         27&         29&         20&         22&         10&         10&          3&          3\\
     1e-7 &      1e-4 &       0.46 &    9.6e-3 &      0.016&    1.3e2&    1.4e2&    6.3e1&    6.5e1&    1.4e1&    1.4e1&    6.0e-1&    6.1e-1 &         43&         45&         48&         49&         57&         58&         27&         27\\
     1e-7 &      1e-1 &       1.85 &    1.6e-4 &      0.016&    2.9&    3.2&    8.1e-1&    8.9e-1&    6.6e-2&    7.3e-2&    1.8e-3&    2.0e-3 &       0.06&       0.06&       0.04&       0.04&       0.02&       0.02&       0.01&       0.01\\
     1e-7 &      1e-2 &       1.85 &    1.3e-3 &      0.016&    2.8e1&    3.1e1&    7.8&    8.6&    6.4e-1&    7.0e-1&    1.7e-2&    1.9e-2 &       0.56&       0.62&       0.37&       0.41&       0.16&       0.18&       0.05&       0.05\\
     1e-7 &      1e-3 &       1.85 &    1.0e-2 &      0.016&    2.3e2&    2.6e2&    6.8e1&    7.6e1&    5.8&    6.4&    1.6e-1&    1.8e-1 &       4.74&       5.28&       3.24&       3.60&       1.47&       1.63&       0.45&       0.50\\
     1e-7 &      1e-4 &       1.85 &    6.7e-2 &      0.016&    8.0e2&    8.8e2&    3.6e2&    3.9e2&    4.4e1&    4.7e1&    1.4&    1.5 &      16.24&      17.94&      17.26&      18.72&      11.02&      11.91&       3.93&       4.21\\
  \hline
    1e-8 &      1e-1 &       0.12 &    1.4e-6 &      0.002&    5.4e-2&    7.0e-2&    1.5e-2&    2.0e-2&    1.2e-3&    1.6e-3&    3.3e-5&    4.3e-5 &       0.28&       0.36&       0.19&       0.24&       0.08&       0.10&       0.02&       0.03\\
     1e-8 &      1e-2 &       0.12 &    8.7e-6 &      0.002&    5.4e-1&    6.2e-1&    1.5e-1&    1.7e-1&    1.2e-2&    1.4e-2&    3.3e-4&    3.8e-4 &       2.78&       3.20&       1.82&       2.09&       0.79&       0.91&       0.24&       0.27\\
     1e-8 &      1e-3 &       0.12 &    5.5e-5 &      0.002&    4.7&    4.9&    1.4&    1.4&    1.2e-1&    1.2e-1&    3.3e-3&    3.4e-3 &      24.46&      25.37&      16.98&      17.58&       7.78&       8.04&       2.35&       2.43\\
     1e-8 &      1e-4 &       0.12 &    3.5e-4 &      0.002&    1.3e1&    1.3e1&    6.2&    6.2&    9.2e-1&    9.3e-1&    3.1e-2&    3.1e-2 &      67.51&      68.27&      75.07&      75.62&      59.84&      60.27&      22.51&      22.65\\
     1e-8 &      1e-1 &       0.46 &    7.0e-6 &      0.002&    1.4e-1&    1.5e-1&    3.9e-2&    4.3e-2&    3.1e-3&    3.5e-3&    8.5e-5&    9.5e-5 &       0.04&       0.05&       0.03&       0.03&       0.01&       0.01&       0.00&       0.00\\
     1e-8 &      1e-2 &       0.46 &    5.6e-5 &      0.002&    1.3&    1.5&    3.7e-1&    4.1e-1&    3.0e-2&    3.4e-2&    8.0e-4&    9.1e-4 &       0.42&       0.48&       0.28&       0.31&       0.12&       0.14&       0.04&       0.04\\
     1e-8 &      1e-3 &       0.46 &    3.8e-4 &      0.002&    1.1e1&    1.3e1&    3.3&    3.7&    2.7e-1&    3.1e-1&    7.4e-3&    8.3e-3 &       3.70&       4.18&       2.48&       2.80&       1.10&       1.24&       0.33&       0.37\\
     1e-8 &      1e-4 &       0.46 &    2.4e-3 &      0.002&    6.9e1&    7.2e1&    2.4e1&    2.5e1&    2.3&    2.4&    6.9e-2&    7.2e-2 &      22.25&      23.37&      17.88&      18.76&       9.39&       9.80&       3.10&       3.23\\
     1e-8 &      1e-1 &       1.85 &    1.9e-5 &      0.002&    3.1e-1&    3.2e-1&    8.6e-2&    9.1e-2&    7.0e-3&    7.4e-3&    1.9e-4&    2.0e-4 &       0.01&       0.01&       0.00&       0.00&       0.00&       0.00&       0.00&       0.00\\
     1e-8 &      1e-2 &       1.85 &    1.8e-4 &      0.002&    3.0&    3.2&    8.4e-1&    8.9e-1&    6.8e-2&    7.2e-2&    1.8e-3&    1.9e-3 &       0.06&       0.06&       0.04&       0.04&       0.02&       0.02&       0.01&       0.01\\
     1e-8 &      1e-3 &       1.85 &    1.6e-3 &      0.002&    2.8e1&    3.0e1&    8.0&    8.4&    6.6e-1&    6.9e-1&    1.8e-2&    1.9e-2 &       0.58&       0.61&       0.38&       0.40&       0.17&       0.18&       0.05&       0.05\\
     1e-8 &      1e-4 &       1.85 &    1.3e-2 &      0.002&    2.3e2&    2.4e2&    6.9e1&    7.2e1&    6.1&    6.3&    1.7e-1&    1.8e-1 &       4.72&       4.89&       3.30&       3.42&       1.53&       1.59&       0.48&       0.49\\
\hline
\end{tabular}
\end{center}
\end{table*}

\begin{table*}
\tiny
\begin{center}
\caption{Irradiated Discs.  The values of $10^{-5}$, $10^{-4}$, $10^{-3}$ in the third row refer to cases with $L_{irr}=10^{-5}L_{\odot}$, $10^{-4}L_{\odot}$, $10^{-3}L_{\odot}$. \label{tab2}}
\setlength\tabcolsep{2.5pt}
\begin{tabular}{ccc|cc|ccc|ccc|ccc|ccc|ccc|ccc}
\hline\hline
$M_{p}\dot{M_{p}}$&$\alpha$&$R_{out}$ & \multicolumn{2}{|c}{$M_{disc}$} & \multicolumn{3}{|c}{F(0.85mm)}  & \multicolumn{3}{|c}{F(1.3mm)} & \multicolumn{3}{|c}{F(3mm)}  &  \multicolumn{3}{|c}{$\overline{T_{b}}(0.85mm)$} & \multicolumn{3}{|c}{$\overline{T_{b}}(1.3mm)$} & \multicolumn{3}{|c}{$\overline{T_{b}}(3mm)$} \\
M$_{J}^2$/yr&&AU& $M_{J}$ & $M_{J}$  & $\mu$Jy & $\mu$Jy& $\mu$Jy & $\mu$Jy & $\mu$Jy & $\mu$Jy & $\mu$Jy & $\mu$Jy & $\mu$Jy & K & K& K& K & K& K & K & K & K\\
& & & $10^{-3}$ & $10^{-5}$  & $10^{-3}$ & $10^{-4}$  & $10^{-5}$ &  $10^{-3}$ & $10^{-4}$  & $10^{-5}$ &  $10^{-3}$ & $10^{-4}$  & $10^{-5}$ &  $10^{-3}$ & $10^{-4}$  & $10^{-5}$ & $10^{-3}$ & $10^{-4}$  & $10^{-5}$ & $10^{-3}$ & $10^{-4}$  & $10^{-5}$\\
\hline
\multicolumn{2}{|c}{$M_{p}=1 M_{J}$}\\
   \hline
     1e-5 &      1e-1 &       0.12 &    8.7e-5 &    8.7e-5&    4.3e1&    4.2e1&    4.2e1&    1.3e1&    1.3e1&    1.3e1&    1.2&    1.2&    1.2 &        223&        220&        219&        161&        159&        159&         77&         76&         76\\
     1e-5 &      1e-2 &       0.12 &    5.5e-4 &    5.5e-4&    7.4e1&    7.3e1&    7.3e1&    3.5e1&    3.5e1&    3.5e1&    7.4&    7.3&    7.3 &        382&        380&        380&        424&        422&        422&        476&        475&        475\\
     1e-5 &      1e-3 &       0.12 &    3.5e-3 &    3.5e-3&    7.4e1&    7.3e1&    7.3e1&    3.5e1&    3.5e1&    3.5e1&    8.1&    8.0&    8.0 &        382&        380&        380&        424&        422&        422&        521&        520&        520\\
     1e-5 &      1e-4 &       0.12 &    2.2e-2 &    2.2e-2&    7.4e1&    7.3e1&    7.3e1&    3.5e1&    3.5e1&    3.5e1&    8.1&    8.0&    8.0 &        382&        380&        380&        424&        422&        422&        521&        520&        520\\
     1e-5 &      1e-1 &       0.46 &    6.1e-4 &    6.1e-4&    1.1e2&    1.0e2&    1.0e2&    3.3e1&    3.0e1&    3.0e1&    2.8&    2.6&    2.5 &         37&         34&         33&         25&         23&         23&         11&         10&         10\\
     1e-5 &      1e-2 &       0.46 &    3.8e-3 &    3.8e-3&    4.3e2&    4.2e2&    4.2e2&    1.9e2&    1.8e2&    1.8e2&    2.1e1&    2.1e1&    2.1e1 &        140&        136&        136&        143&        140&        140&         87&         85&         85\\
     1e-5 &      1e-3 &       0.46 &    2.4e-2 &    2.4e-2&    4.3e2&    4.2e2&    4.2e2&    2.0e2&    2.0e2&    2.0e2&    4.6e1&    4.6e1&    4.6e1 &        140&        136&        136&        154&        152&        151&        188&        187&        186\\
     1e-5 &      1e-4 &       0.46 &    1.5e-1 &    1.5e-1&    4.3e2&    4.2e2&    4.2e2&    2.0e2&    2.0e2&    2.0e2&    4.6e1&    4.6e1&    4.6e1 &        140&        136&        136&        154&        152&        151&        188&        187&        186\\
     1e-5 &      1e-1 &       1.85 &    3.6e-3 &    4.2e-3&    2.9e2&    2.4e2&    2.3e2&    8.1e1&    6.9e1&    6.5e1&    6.7&    5.8&    5.4 &       5.79&       4.95&       4.62&       3.85&       3.30&       3.08&       1.70&       1.46&       1.36\\
     1e-5 &      1e-2 &       1.85 &    2.7e-2 &    2.7e-2&    1.9e3&    1.7e3&    1.6e3&    6.0e2&    5.3e2&    5.2e2&    5.5e1&    4.9e1&    4.9e1 &         38&         34&         33&         28&         25&         25&         14&         13&         12\\
     1e-5 &      1e-3 &       1.85 &    1.7e-1 &    1.7e-1&    2.6e3&    2.4e3&    2.4e3&    1.2e3&    1.1e3&    1.1e3&    2.7e2&    2.6e2&    2.6e2 &         53&         49&         48&         57&         54&         54&         68&         67&         66\\
     1e-5 &      1e-4 &       1.85 &    1.1 &    1.1&    2.6e3&    2.4e3&    2.4e3&    1.2e3&    1.1e3&    1.1e3&    2.7e2&    2.6e2&    2.6e2 &         53&         49&         48&         57&         54&         54&         68&         67&         66\\
     \hline
    1e-6 &      1e-1 &       0.12 &    2.1e-5 &    2.2e-5&    6.5&    5.5&    5.2&    1.9&    1.6&    1.5&    1.5e-1&    1.3e-1&    1.2e-1 &         34&         28&         27&         22&         19&         18&         10&          8&          8\\
     1e-6 &      1e-2 &       0.12 &    1.4e-4 &    1.4e-4&    3.8e1&    3.5e1&    3.5e1&    1.3e1&    1.2e1&    1.2e1&    1.2&    1.1&    1.1 &        199&        183&        182&        155&        143&        142&         79&         74&         73\\
     1e-6 &      1e-3 &       0.12 &    8.7e-4 &    8.7e-4&    4.4e1&    4.1e1&    4.1e1&    2.0e1&    2.0e1&    2.0e1&    4.6&    4.5&    4.5 &        227&        215&        214&        248&        238&        237&        298&        293&        293\\
     1e-6 &      1e-4 &       0.12 &    5.5e-3 &    5.5e-3&    4.4e1&    4.1e1&    4.1e1&    2.0e1&    2.0e1&    2.0e1&    4.6&    4.5&    4.5 &        227&        215&        214&        248&        238&        237&        298&        293&        293\\
     1e-6 &      1e-1 &       0.46 &    9.6e-5 &    1.5e-4&    1.5e1&    1.4e1&    1.2e1&    4.2&    3.9&    3.3&    3.4e-1&    3.2e-1&    2.7e-1 &       4.88&       4.46&       3.81&       3.21&       2.94&       2.51&       1.39&       1.27&       1.09\\
     1e-6 &      1e-2 &       0.46 &    8.7e-4 &    9.6e-4&    1.2e2&    1.0e2&    9.6e1&    3.6e1&    3.0e1&    2.9e1&    3.1&    2.7&    2.5 &         40&         33&         31&         28&         23&         22&         13&         11&         10\\
     1e-6 &      1e-3 &       0.46 &    6.1e-3 &    6.1e-3&    2.8e2&    2.4e2&    2.4e2&    1.3e2&    1.1e2&    1.1e2&    2.0e1&    1.9e1&    1.8e1 &         91&         79&         77&         97&         87&         85&         83&         75&         74\\
     1e-6 &      1e-4 &       0.46 &    3.8e-2 &    3.8e-2&    2.8e2&    2.4e2&    2.4e2&    1.3e2&    1.1e2&    1.1e2&    2.8e1&    2.6e1&    2.6e1 &         91&         79&         77&         97&         87&         85&        112&        106&        105\\
     1e-6 &      1e-1 &       1.85 &    4.0e-4 &    9.6e-4&    3.2e1&    3.1e1&    2.8e1&    9.0&    8.6&    8.0&    7.3e-1&    7.0e-1&    6.5e-1 &       0.65&       0.63&       0.58&       0.43&       0.41&       0.38&       0.18&       0.18&       0.16\\
     1e-6 &      1e-2 &       1.85 &    3.9e-3 &    6.7e-3&    2.9e2&    2.7e2&    2.4e2&    8.4e1&    7.8e1&    6.8e1&    7.0&    6.5&    5.7 &       5.95&       5.52&       4.78&       3.99&       3.70&       3.22&       1.78&       1.65&       1.44\\
     1e-6 &      1e-3 &       1.85 &    3.6e-2 &    4.2e-2&    1.7e3&    1.4e3&    1.3e3&    6.0e2&    5.0e2&    4.6e2&    6.0e1&    5.0e1&    4.7e1 &         35&         29&         27&         28&         24&         22&         15&         13&         12\\
     1e-6 &      1e-4 &       1.85 &    2.7e-1 &    2.7e-1&    2.0e3&    1.5e3&    1.4e3&    8.8e2&    6.8e2&    6.4e2&    1.8e2&    1.5e2&    1.5e2 &         40&         30&         28&         42&         32&         31&         45&         38&         37\\
     \hline
     1e-7 &      1e-1 &       0.12 &    2.5e-6 &    5.5e-6&    7.6e-1&    7.1e-1&    6.2e-1&    2.1e-1&    2.0e-1&    1.7e-1&    1.7e-2&    1.6e-2&    1.4e-2 &       3.93&       3.70&       3.20&       2.57&       2.42&       2.09&       1.11&       1.05&       0.91\\
     1e-7 &      1e-2 &       0.12 &    2.4e-5 &    3.5e-5&    6.9&    6.1&    5.2&    2.0&    1.8&    1.5&    1.6e-1&    1.5e-1&    1.3e-1&      35.77&      31.76&      26.98&      24.06&      21.39&      18.26&      10.64&       9.47&       8.11\\
     1e-7 &      1e-3 &       0.12 &    2.1e-4 &    2.2e-4&    3.2e1&    2.5e1&    2.3e1&    1.2e1&    1.0e1&    9.6&    1.3&    1.1&    1.1 &     166&     128&     121&     151&     124&     117&      86.81&      71.67&      68.18\\
     1e-7 &      1e-4 &       0.12 &    1.4e-3 &    1.4e-3&    3.2e1&    2.5e1&    2.3e1&    1.4e1&    1.1e1&    1.1e1&    2.9&    2.6&    2.5 &     166&     128&     121&     172&     139&     134&     190&     168&     165\\
     1e-7 &      1e-1 &       0.46 &    1.0e-5 &    2.9e-5&    1.6&    1.6&    1.5&    4.5e-1&    4.4e-1&    4.1e-1&    3.6e-2&    3.5e-2&    3.4e-2 &       0.52&       0.51&       0.48&       0.34&       0.33&       0.31&       0.15&       0.14&       0.14\\
     1e-7 &      1e-2 &       0.46 &    9.9e-5 &    2.4e-4&    1.5e1&    1.5e1&    1.3e1&    4.3&    4.1&    3.7&    3.6e-1&    3.4e-1&    3.1e-1 &       4.98&       4.76&       4.26&       3.30&       3.15&       2.82&       1.44&       1.38&       1.24\\
     1e-7 &      1e-3 &       0.46 &    9.6e-4 &    1.5e-3&    1.2e2&    1.1e2&    8.9e1&    3.6e1&    3.3e1&    2.8e1&    3.3&    3.0&    2.5 &      37.93&      34.79&      28.77&      27.46&      25.24&      21.06&      13.25&      12.07&      10.23\\
     1e-7 &      1e-4 &       0.46 &    8.7e-3 &    9.6e-3&    2.4e2&    1.6e2&    1.4e2&    1.0e2&    7.2e1&    6.4e1&    1.9e1&    1.5e1&    1.4e1 &      78.11&      51.48&      44.20&      79.18&      54.58&      48.71&      75.68&      61.48&      57.49\\
     1e-7 &      1e-1 &       1.85 &    4.0e-5 &    1.2e-4&    3.3&    3.3&    3.2&    9.2e-1&    9.1e-1&    8.9e-1&    7.5e-2&    7.4e-2&    7.2e-2 &       0.07&       0.07&       0.06&       0.04&       0.04&       0.04&       0.02&       0.02&       0.02\\
     1e-7 &      1e-2 &       1.85 &    4.0e-4 &    1.1e-3&    3.2e1&    3.2e1&    3.0e1&    9.1&    8.9&    8.5&    7.4e-1&    7.3e-1&    6.9e-1 &       0.66&       0.64&       0.61&       0.43&       0.42&       0.40&       0.19&       0.18&       0.18\\
     1e-7 &      1e-3 &       1.85 &    4.0e-3 &    9.6e-3&    2.9e2&    2.8e2&    2.5e2&    8.4e1&    8.1e1&    7.4e1&    7.1&    6.9&    6.3 &       5.81&       5.64&       5.17&       3.97&       3.84&       3.52&       1.80&       1.74&       1.60\\
     1e-7 &      1e-4 &       1.85 &    3.9e-2 &    6.7e-2&    1.6e3&    1.1e3&    8.6e2&    5.6e2&    4.6e2&    3.8e2&    5.7e1&    5.4e1&    4.6e1 &      33.32&      23.00&      17.39&      26.39&      21.69&      18.24&      14.50&      13.60&      11.62\\
     \hline
     1e-8 &      1e-1 &       0.12 &    2.5e-7 &    7.7e-7&    7.9e-2&    7.7e-2&    7.4e-2&    2.2e-2&    2.2e-2&    2.1e-2&    1.8e-3&    1.8e-3&    1.7e-3 &       0.41&       0.40&       0.39&       0.27&       0.26&       0.25&       0.12&       0.11&       0.11\\
     1e-8 &      1e-2 &       0.12 &    2.5e-6 &    7.2e-6&    7.8e-1&    7.5e-1&    7.0e-1&    2.2e-1&    2.1e-1&    2.0e-1&    1.8e-2&    1.7e-2&    1.6e-2 &       4.03&       3.89&       3.63&       2.64&       2.54&       2.37&       1.14&       1.10&       1.03\\
     1e-8 &      1e-3 &       0.12 &    2.5e-5 &    5.5e-5&    6.7&    6.4&    5.5&    2.0&    1.9&    1.6&    1.7e-1&    1.6e-1&    1.4e-1 &      35.02&      33.39&      28.68&      24.15&      22.97&      19.76&      11.03&      10.40&       8.99\\
     1e-8 &      1e-4 &       0.12 &    2.4e-4 &    3.5e-4&    3.0e1&    1.8e1&    1.4e1&    1.1e1&    8.0&    6.4&    1.3&    1.2&    9.6e-1 &     154&      94 &      72&     137&      97&      78.37&      83.52&      74.90&      62.33\\
     1e-8 &      1e-1 &       0.46 &    1.0e-6 &    3.1e-6&    1.6e-1&    1.6e-1&    1.6e-1&    4.6e-2&    4.5e-2&    4.4e-2&    3.7e-3&    3.7e-3&    3.6e-3 &       0.05&       0.05&       0.05&       0.03&       0.03&       0.03&       0.02&       0.01&       0.01\\
     1e-8 &      1e-2 &       0.46 &    1.0e-5 &    3.1e-5&    1.6&    1.6&    1.5&    4.5e-1&    4.5e-1&    4.3e-1&    3.7e-2&    3.6e-2&    3.5e-2 &       0.53&       0.52&       0.50&       0.34&       0.34&       0.33&       0.15&       0.15&       0.14\\
     1e-8 &      1e-3 &       0.46 &    1.0e-4 &    2.9e-4&    1.5e1&    1.5e1&    1.4e1&    4.3&    4.3&    4.0&    3.6e-1&    3.5e-1&    3.3e-1 &       4.93&       4.83&       4.59&       3.30&       3.23&       3.06&       1.47&       1.43&       1.35\\
     1e-8 &      1e-4 &       0.46 &    9.9e-4 &    2.4e-3&    1.1e2&    9.8e1&    8.3e1&    3.5e1&    3.2e1&    2.9e1&    3.2&    3.1&    2.8 &      36.98&      31.74&      27.11&      26.48&      24.14&      21.70&      12.99&      12.54&      11.17\\
     1e-8 &      1e-1 &       1.85 &    4.0e-6 &    1.2e-5&    3.3e-1&    3.3e-1&    3.3e-1&    9.3e-2&    9.3e-2&    9.2e-2&    7.6e-3&    7.5e-3&    7.5e-3 &       0.01&       0.01&       0.01&       0.00&       0.00&       0.00&       0.00&       0.00&       0.00\\
     1e-8 &      1e-2 &       1.85 &    4.0e-5 &    1.2e-4&    3.3&    3.3&    3.2&    9.3e-1&    9.2e-1&    9.1e-1&    7.6e-2&    7.5e-2&    7.4e-2 &       0.07&       0.07&       0.07&       0.04&       0.04&       0.04&       0.02&       0.02&       0.02\\
     1e-8 &      1e-3 &       1.85 &    4.0e-4 &    1.2e-3&    3.2e1&    3.2e1&    3.1e1&    9.1&    9.0&    8.8&    7.5e-1&    7.4e-1&    7.2e-1 &       0.65&       0.65&       0.63&       0.43&       0.43&       0.42&       0.19&       0.19&       0.18\\
     1e-8 &      1e-4 &       1.85 &    4.0e-3 &    1.1e-2&    2.8e2&    2.7e2&    2.5e2&    8.2e1&    7.9e1&    7.6e1&    7.1&    7.0&    6.7 &       5.75&       5.42&       5.16&       3.90&       3.76&       3.63&       1.79&       1.76&       1.68\\
\hline
\multicolumn{22}{l}{$M_{p}=10 M_{J}$, $R_{out}$ of 0.12, 0.46, 1.85 AU corresponds to CPDs around a 10 $M_{J}$ planet at 2.3, 9.3, 37 AU from the central solar mass star.}\\
 \hline
     1e-5 &      1e-1 &       0.12 &    3.5e-5 &    3.5e-5 &    1.6e1&    1.6e1&    1.6e1&    4.7&    4.6&    4.6&    4.0e-1&    3.8e-1&    3.8e-1 &         85&         83&         82&         58&         56&         56&         26&         25&         25\\
     1e-5 &      1e-2 &       0.12 &    2.2e-4 &    2.2e-4 &    7.4e1&    7.3e1&    7.3e1&    3.0e1&    3.0e1&    3.0e1&    3.3&    3.3&    3.3 &        382&        380&        380&        370&        367&        367&        216&        214&        214\\
     1e-5 &      1e-3 &       0.12 &    1.4e-3 &    1.4e-3 &    7.4e1&    7.3e1&    7.3e1&    3.5e1&    3.5e1&    3.5e1&    8.1&    8.0&    8.0 &        382&        380&        380&        424&        422&        422&        521&        520&        520\\
     1e-5 &      1e-4 &       0.12 &    8.7e-3 &    8.7e-3 &    7.4e1&    7.3e1&    7.3e1&    3.5e1&    3.5e1&    3.5e1&    8.1&    8.0&    8.0 &        382&        380&        380&        424&        422&        422&        521&        520&        520\\
     1e-5 &      1e-1 &       0.46 &    2.4e-4 &    2.4e-4 &    4.1e1&    3.6e1&    3.5e1&    1.2e1&    1.0e1&    9.9&    9.6e-1&    8.4e-1&    8.2e-1 &         13&         12&         11&          9&          8&          8&          4&          3&          3\\
     1e-5 &      1e-2 &       0.46 &    1.5e-3 &    1.5e-3 &    2.8e2&    2.7e2&    2.7e2&    8.8e1&    8.4e1&    8.3e1&    8.0&    7.7&    7.6  &         91&         87&         86&         66&         64&         63&         32&         31&         31\\
     1e-5 &      1e-3 &       0.46 &    9.6e-3 &    9.6e-3 &    4.3e2&    4.2e2&    4.2e2&    2.0e2&    2.0e2&    2.0e2&    4.5e1&    4.5e1&    4.5e1 &        140&        136&        136&        154&        152&        151&        182&        180&        180\\
     1e-5 &      1e-4 &       0.46 &    6.1e-2 &    6.1e-2 &    4.3e2&    4.2e2&    4.2e2&    2.0e2&    2.0e2&    2.0e2&    4.6e1&    4.6e1&    4.6e1 &        140&        136&        136&        154&        152&        151&        188&        187&        186\\
     1e-5 &      1e-1 &       1.85 &    1.2e-3 &    1.7e-3 &    9.6e1&    8.6e1&    7.6e1&    2.7e1&    2.4e1&    2.1e1&    2.2&    2.0&    1.8 &       1.94&       1.74&       1.54&       1.28&       1.15&       1.02&       0.56&       0.50&       0.44\\
     1e-5 &      1e-2 &       1.85 &    1.0e-2 &    1.1e-2 &    8.0e2&    6.8e2&    6.5e2&    2.3e2&    2.0e2&    1.9e2&    2.0e1&    1.7e1&    1.6e1 &         16&         14&         13&         11&          9&          9&          5&          4&          4\\
     1e-5 &      1e-3 &       1.85 &    6.7e-2 &    6.7e-2 &    2.6e3&    2.4e3&    2.4e3&    1.2e3&    1.1e3&    1.1e3&    1.4e2&    1.3e2&    1.3e2 &         53&         49&         48&         56&         52&         52&         36&         34&         33\\
     1e-5 &      1e-4 &       1.85 &    4.2e-1 &    4.2e-1 &    2.6e3&    2.4e3&    2.4e3&    1.2e3&    1.1e3&    1.1e3&    2.7e2&    2.6e2&    2.6e2 &         53&         49&         48&         57&         54&         54&         68&         67&         66\\
     \hline
     1e-6 &      1e-1 &       0.12 &    7.2e-6 &    8.7e-6 &    2.2&    1.9&    1.7&    6.2e-1&    5.2e-1&    4.8e-1&    5.0e-2&    4.3e-2&    3.9e-2 &         11&         10&          9&          8&          6&          6&          3&          3&          3\\
     1e-6 &      1e-2 &       0.12 &    5.5e-5 &    5.5e-5 &    1.7e1&    1.5e1&    1.5e1&    5.1&    4.5&    4.4&    4.4e-1&    3.9e-1&    3.8e-1 &         91&         79&         77&         62&         55&         54&         28&         25&         25\\
     1e-6 &      1e-3 &       0.12 &    3.5e-4 &    3.5e-4 &    4.4e1&    4.1e1&    4.1e1&    2.0e1&    2.0e1&    2.0e1&    3.0&    2.9&    2.9 &        227&        215&        214&        248&        238&        237&        197&        190&        189\\
     1e-6 &      1e-4 &       0.12 &    2.2e-3 &    2.2e-3 &    4.4e1&    4.1e1&    4.1e1&    2.0e1&    2.0e1&    2.0e1&    4.6&    4.5&    4.5 &        227&        215&        214&        248&        238&        237&        298&        293&        293\\
     1e-6 &      1e-1 &       0.46 &    3.1e-5 &    6.1e-5 &    4.9&    4.6&    3.9&    1.4&    1.3&    1.1&    1.1e-1&    1.0e-1&    8.9e-2 &       1.59&       1.49&       1.28&       1.04&       0.97&       0.83&       0.45&       0.42&       0.36\\
     1e-6 &      1e-2 &       0.46 &    2.9e-4 &    3.8e-4 &    4.5e1&    3.9e1&    3.5e1&    1.3e1&    1.1e1&    9.9&    1.1&    9.2e-1&    8.3e-1 &         15&         13&         11&         10&          8&          8&          4&          4&          3\\
     1e-6 &      1e-3 &       0.46 &    2.4e-3 &    2.4e-3 &    2.6e2&    2.2e2&    2.1e2&    9.0e1&    7.6e1&    7.3e1&    8.7&    7.5&    7.3 &         86&         72&         69&         68&         58&         56&         35&         30&         29\\
     1e-6 &      1e-4 &       0.46 &    1.5e-2 &    1.5e-2 &    2.8e2&    2.4e2&    2.4e2&    1.3e2&    1.1e2&    1.1e2&    2.8e1&    2.6e1&    2.6e1 &         91&         79&         77&         97&         87&         85&        112&        106&        105\\
     1e-6 &      1e-1 &       1.85 &    1.3e-4 &    3.3e-4 &    1.0e1&    1.0e1&    9.4&    2.9&    2.8&    2.6&    2.3e-1&    2.3e-1&    2.1e-1 &       0.21&       0.20&       0.19&       0.14&       0.13&       0.12&       0.06&       0.06&       0.05\\
     1e-6 &      1e-2 &       1.85 &    1.2e-3 &    2.6e-3 &    9.8e1&    9.3e1&    8.4e1&    2.8e1&    2.6e1&    2.4e1&    2.3&    2.2&    1.9 &       2.00&       1.90&       1.70&       1.32&       1.25&       1.12&       0.58&       0.55&       0.49\\
     1e-6 &      1e-3 &       1.85 &    1.2e-2 &    1.7e-2 &    8.1e2&    7.2e2&    6.2e2&    2.4e2&    2.2e2&    1.9e2&    2.1e1&    1.9e1&    1.7e1 &         16&         15&         13&         11&         10&          9&          5&          5&          4\\
     1e-6 &      1e-4 &       1.85 &    1.0e-1 &    1.1e-1 &    2.0e3&    1.5e3&    1.4e3&    8.8e2&    6.8e2&    6.4e2&    1.4e2&    1.2e2&    1.1e2 &         40&         30&         28&         42&         32&         31&         36&         30&         29\\
     \hline
     1e-7 &      1e-1 &       0.12 &    7.9e-7 &    2.1e-6 &    2.4e-1&    2.3e-1&    2.1e-1&    6.8e-2&    6.5e-2&    5.9e-2&    5.5e-3&    5.3e-3 &    4.8e-3 &       1.26&       1.20&       1.09&       0.82&       0.79&       0.71&       0.36&       0.34&       0.31\\
     1e-7 &      1e-2 &       0.12 &    7.7e-6 &    1.4e-5 &    2.3&    2.2&    1.8&    6.5e-1&    6.0e-1&    5.1e-1&    5.3e-2&    4.9e-2&    4.1e-2 &      12.12&      11.19&       9.37&       7.96&       7.35&       6.16&       3.45&       3.18&       2.67\\
     1e-7 &      1e-3 &       0.12 &    7.2e-5 &    8.7e-5 &    1.8e1&    1.5e1&    1.4e1&    5.6&    4.6&    4.2&    4.9e-1&    4.1e-1&    3.8e-1 &      94.07&      78.53&      70.56&      67.48&      56.38&      51.00&      31.73&      26.74&      24.39\\
     1e-7 &      1e-4 &       0.12 &    5.5e-4 &    5.5e-4 &    3.2e1&    2.5e1&    2.3e1&    1.4e1&    1.1e1&    1.1e1&    2.8&    2.4&    2.3 &     166&     128&     121&     172&     139&     134&     180&     154&     151\\
     1e-7 &      1e-1 &       0.46 &    3.2e-6 &    9.6e-6 &    5.1e-1&    5.0e-1&    4.8e-1&    1.4e-1&    1.4e-1&    1.3e-1&    1.2e-2&    1.1e-2&    1.1e-2 &       0.17&       0.16&       0.16&       0.11&       0.11&       0.10&       0.05&       0.05&       0.04\\
     1e-7 &      1e-2 &       0.46 &    3.2e-5 &    8.6e-5 &    5.0&    4.8&    4.5&    1.4&    1.4&    1.3&    1.1e-1&    1.1e-1&    1.0e-1 &       1.63&       1.57&       1.46&       1.07&       1.03&       0.96&       0.46&       0.45&       0.41\\
     1e-7 &      1e-3 &       0.46 &    3.1e-4 &    6.1e-4 &    4.5e1&    4.2e1&    3.6e1&    1.3e1&    1.2e1&    1.0e1&    1.1&    1.0&    8.8e-1 &      14.57&      13.74&      11.62&       9.90&       9.31&       7.91&       4.45&       4.18&       3.57\\
     1e-7 &      1e-4 &       0.46 &    2.9e-3 &    3.8e-3 &    2.3e2&    1.6e2&    1.4e2&    8.2e1&    6.8e1&    6.0e1&    9.0&    7.7&    6.8 &      74.50&      51.48&      44.20&      62.40&      51.85&      45.30&      36.29&      31.20&      27.52\\
     1e-7 &      1e-1 &       1.85 &    1.3e-5 &    3.7e-5 &    1.0&    1.0&    1.0&    2.9e-1&    2.9e-1&    2.8e-1&    2.4e-2&    2.4e-2&    2.3e-2 &       0.02&       0.02&       0.02&       0.01&       0.01&       0.01&       0.01&       0.01&       0.01\\
     1e-7 &      1e-2 &       1.85 &    1.3e-4 &    3.6e-4 &    1.0e1&    1.0e1&    9.9&    2.9&    2.9&    2.8&    2.4e-1&    2.3e-1&    2.2e-1 &       0.21&       0.21&       0.20&       0.14&       0.14&       0.13&       0.06&       0.06&       0.06\\
     1e-7 &      1e-3 &       1.85 &    1.3e-3 &    3.3e-3 &    9.9e1&    9.6e1&    9.0e1&    2.8e1&    2.7e1&    2.6e1&    2.3&    2.3&    2.1 &       2.00&       1.95&       1.83&       1.33&       1.29&       1.22&       0.59&       0.57&       0.54\\
     1e-7 &      1e-4 &       1.85 &    1.2e-2 &    2.6e-2 &    7.6e2&    7.0e2&    6.2e2&    2.3e2&    2.2e2&    2.0e2&    2.1e1&    2.0e1&    1.8e1 &      15.53&      14.12&      12.64&      11.01&      10.48&       9.32&       5.35&       5.09&       4.53\\
     \hline
     1e-8 &      1e-1 &       0.12 &    7.9e-8 &    2.5e-7 &    2.5e-2&    2.5e-2&    2.4e-2&    7.0e-3&    6.9e-3&    6.7e-3&    5.7e-4&    5.6e-4&    5.4e-4 &       0.13&       0.13&       0.12&       0.08&       0.08&       0.08&       0.04&       0.04&       0.04\\
     1e-8 &      1e-2 &       0.12 &    7.9e-7 &    2.4e-6 &    2.5e-1&    2.4e-1&    2.3e-1&    6.9e-2&    6.7e-2&    6.4e-2&    5.6e-3&    5.5e-3&    5.2e-3 &       1.29&       1.25&       1.19&       0.84&       0.82&       0.78&       0.36&       0.35&       0.34\\
     1e-8 &      1e-3 &       0.12 &    7.9e-6 &    2.1e-5 &    2.4&    2.3&    2.1&    6.7e-1&    6.4e-1&    5.9e-1&    5.5e-2&    5.3e-2&    4.8e-2 &      12.29&      11.81&      10.73&       8.18&       7.83&       7.11&       3.57&       3.41&       3.10\\
     1e-8 &      1e-4 &       0.12 &    7.7e-5 &    1.4e-4 &    1.7e1&    1.5e1&    1.2e1&    5.3&    4.9&    4.0&    5.0e-1&    4.6e-1&    3.8e-1 &      88.36&      76.46&      62.84&      64.51&      59.72&      48.97&      32.37&      30.00&      24.90\\
     1e-8 &      1e-1 &       0.46 &    3.2e-7 &    1.0e-6 &    5.2e-2&    5.1e-2&    5.1e-2&    1.4e-2&    1.4e-2&    1.4e-2&    1.2e-3&    1.2e-3&    1.2e-3 &       0.02&       0.02&       0.02&       0.01&       0.01&       0.01&       0.00&       0.00&       0.00\\
     1e-8 &      1e-2 &       0.46 &    3.2e-6 &    9.9e-6 &    5.1e-1&    5.1e-1&    5.0e-1&    1.4e-1&    1.4e-1&    1.4e-1&    1.2e-2&    1.2e-2&    1.1e-2 &       0.17&       0.17&       0.16&       0.11&       0.11&       0.11&       0.05&       0.05&       0.05\\
     1e-8 &      1e-3 &       0.46 &    3.2e-5 &    9.6e-5 &    5.0&    4.9&    4.8&    1.4&    1.4&    1.3&    1.2e-1&    1.1e-1&    1.1e-1 &       1.64&       1.61&       1.54&       1.08&       1.06&       1.02&       0.47&       0.46&       0.44\\
     1e-8 &      1e-4 &       0.46 &    3.2e-4 &    8.6e-4 &    4.4e1&    4.1e1&    3.9e1&    1.3e1&    1.2e1&    1.2e1&    1.1&    1.1&    1.0 &      14.19&      13.47&      12.63&       9.70&       9.42&       8.76&       4.48&       4.34&       4.03\\
     1e-8 &      1e-1 &       1.85 &    1.3e-6 &    3.7e-6 &    1.1e-1&    1.1e-1&    1.0e-1&    2.9e-2&    2.9e-2&    2.9e-2&    2.4e-3&    2.4e-3&    2.4e-3 &       0.00&       0.00&       0.00&       0.00&       0.00&       0.00&       0.00&       0.00&       0.00\\
     1e-8 &      1e-2 &       1.85 &    1.3e-5 &    3.7e-5 &    1.1&    1.0&    1.0&    2.9e-1&    2.9e-1&    2.9e-1&    2.4e-2&    2.4e-2&    2.3e-2 &       0.02&       0.02&       0.02&       0.01&       0.01&       0.01&       0.01&       0.01&       0.01\\
     1e-8 &      1e-3 &       1.85 &    1.3e-4 &    3.7e-4 &    1.0e1&    1.0e1&    1.0e1&    2.9&    2.9&    2.8&    2.4e-1&    2.4e-1&    2.3e-1 &       0.21&       0.21&       0.21&       0.14&       0.14&       0.13&       0.06&       0.06&       0.06\\
     1e-8 &      1e-4 &       1.85 &    1.3e-3 &    3.6e-3 &    9.7e1&    9.5e1&    9.3e1&    2.8e1&    2.7e1&    2.7e1&    2.3&    2.3&    2.2 &       1.98&       1.93&       1.88&       1.32&       1.30&       1.26&       0.59&       0.58&       0.56\\
\hline
\end{tabular}
\end{center}
\end{table*}

Tables 1 and 2 summarize the fluxes and the averaged effective brightness temperatures  for all our cases when the source 
is 140 pc away (e.g. in the Taurus star forming region). The planet mass is assumed to be 1 $M_{J}$ in Table 1, and
1 $M_{J}$ or 10 $M_{J}$ in Table 2.
Table 1 shows the discs in the viscous limit, which means that 
the planet is faint (e.g. it had a ``cold start'') and the viscous heating determines the disc temperature structure. 
For such viscous discs, the disc's effective temperature only
depends on the product of $M_{p}$ and $\dot{M_{p}}$ (Equation \ref{eq:Fv}),
and, when the disc is optically thick at mm, the mm emission 
will only depend on  $M_{p}\dot{M_{p}}$. Thus, we use 
$M_{p}\dot{M_{p}}$, $\alpha$, and $R_{out}$ as three fundamental parameters.
However, when the disc is optically thin or is irradiated by the bright planet, 
the flux will also depend on $M_{p}$ (Equations \ref{eq:Sigma} and \ref{eq:Sigma2}, more discussion
in \S 3.2, and 4.1). 

In Table 1, under each temperature and flux column,
there are two sub-columns:  ``b.'' refers to
the cases which consider the boundary layer irradiation ($L_{irr}=L_{acc}=GM_{p}\dot{M_{p}}/(2R_{in})$),
and ``no b.'' refers to the cases without considering the boundary layer irradiation. 

Table 2 shows the cases
where the planet is bright (e.g. the planet has a ``hot start'') and the irradiation from the central planet to the disc is important for determining the disc temperature.
Three planet luminosities ($L_{irr}=10^{-3}, 10^{-4}, 10^{-5} L_{\odot}$) have been considered.
In Table 2,  we list results with both
$M_{p}=1 M_{J}$ and $M_{p}=10 M_{J}$. Notice that with $M_{p}=10 M_{J}$, the disc with $R_{out}=0.48 AU$ corresponds to
a CPD at 9.3 AU instead of 20 AU. 
 When the irradiation is weak (e.g. $L_{p}=10^{-5} L_{\odot}$), viscous heating dominates
and the flux of CPDs with $M_{p}=1 M_{J}$ in Table 2 is similar to those in Table 1.

\section{Results}
In \S 3.1, we first present   the results without including the irradiation from the planet to the disc (cases in Table 1). 
Then, in \S 3.2, we present the results which consider the planet irradiation (cases in Table 2).

\subsection{Viscous Heating Dominated Discs}

\begin{figure*} 
\centering
\includegraphics[trim=0cm 0cm 0cm 0cm, width=0.9\textwidth]{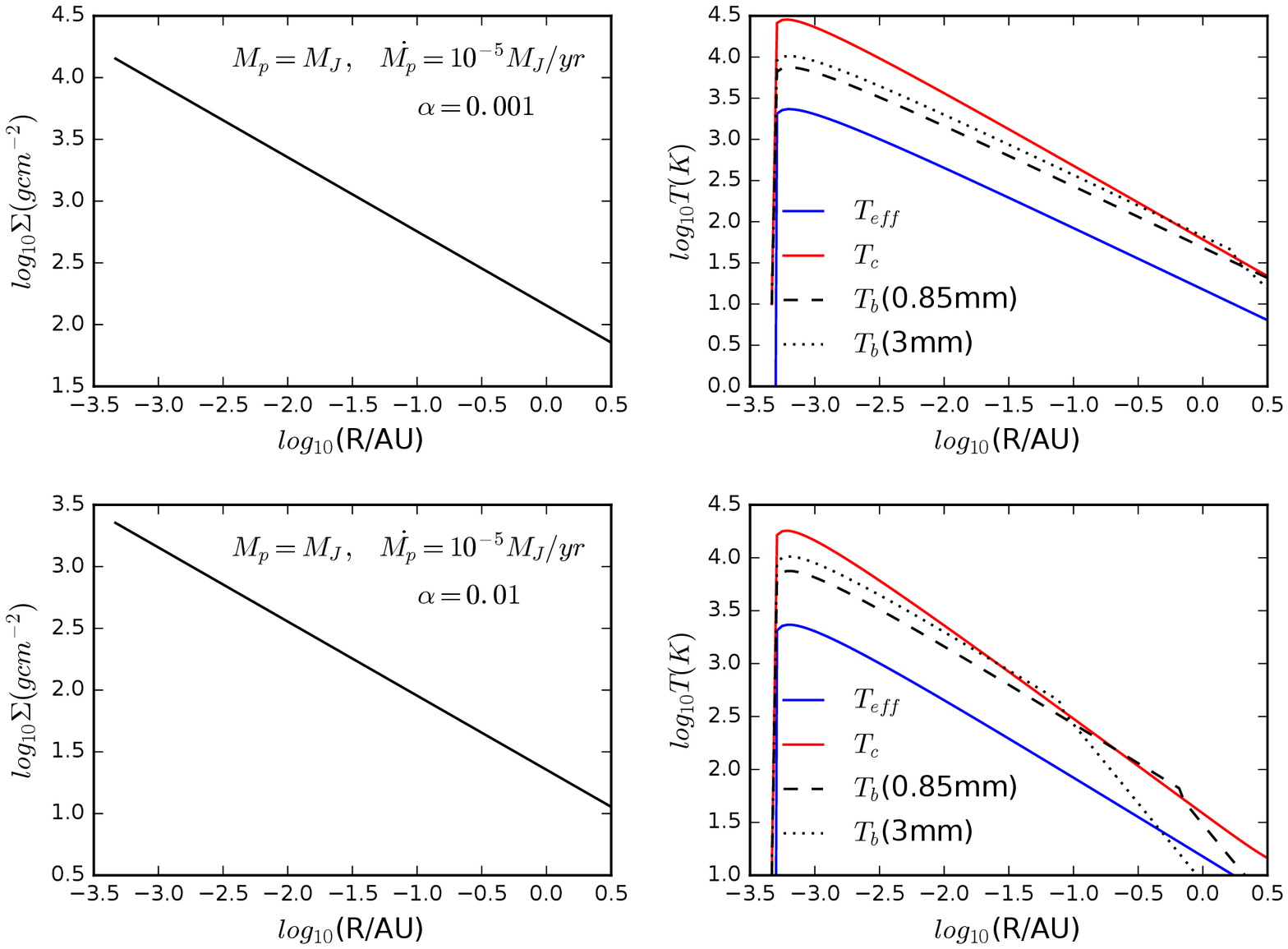} 
\caption{ The disc surface density (left panels) and temperatures (right panels) for viscous heating dominated CPDs with
$\dot{M_{p}}=10^{-5}$ M$_{J}$/yr, $M_{p}=1 M_{J}$, and $\alpha$=0.001 (upper panels) or $\alpha$=0.01 (lower panels).
Disc effective temperature, midplane temperature, effective brightness temperature at 0.85 mm and 3mm are shown in 
the right panels. }
\vspace{-0.1 cm} \label{fig:oned}
\end{figure*}

Without the planet irradiation, the disc thermal structure is solely determined by the viscous accretion.
The disc structure for CPDs with $\dot{M_{p}}=10^{-5}$ M$_{J}$/yr and $M_{p}=1 M_{J}$ is shown in Figure \ref{fig:oned}.
Two cases with different $\alpha$ values ($\alpha=0.01$ and 0.001) are presented. The disc
surface density follows R$^{-3/5}$, as  in Equation \ref{eq:Sigma}. 
With a smaller $\alpha$ (upper panels), 
the disc surface density is higher in order to maintain the same accretion rate. 
Since this disc is almost optically thick at mm within 2 AU,
the effective brightness temperature at mm is always higher than the effective temperature calculated with the Rosseland mean opacity.
With a higher $\alpha$ (lower panels), the surface density is lower
and the outer disc becomes optically thin at mm, so that the brightness
temperature drops off faster there. 

The figure also suggests that the midplane temperature at the very inner disc ($<0.03$ AU or 60 $R_{J}$) can be higher than the dust sublimation
temperature ( $\sim$ 1500 K). Dust will sublimate at the inner disc and the disc will be ionized. The ionized plasma at the inner disc
can also radiate at mm from free-free emission. For example, at 100 GHz our sun's brightness temperature  ($\sim$7000 K) is similar to the temperature of the
photosphere. Thus, we simply assume that the brightness temperature from the free-free emission at the inner disc is the same
as the brightness temperature calculated using dust opacity. Furthermore, as we will show in \S 4.1, most of the radio emission comes from the outer disc, and
changing the inner radius from $R_{J}$ to 100 $R_{J}$ has little effect on the radio emission. 

\begin{figure} 
\centering
\includegraphics[trim=0cm 0cm 0cm 0cm, width=0.5\textwidth]{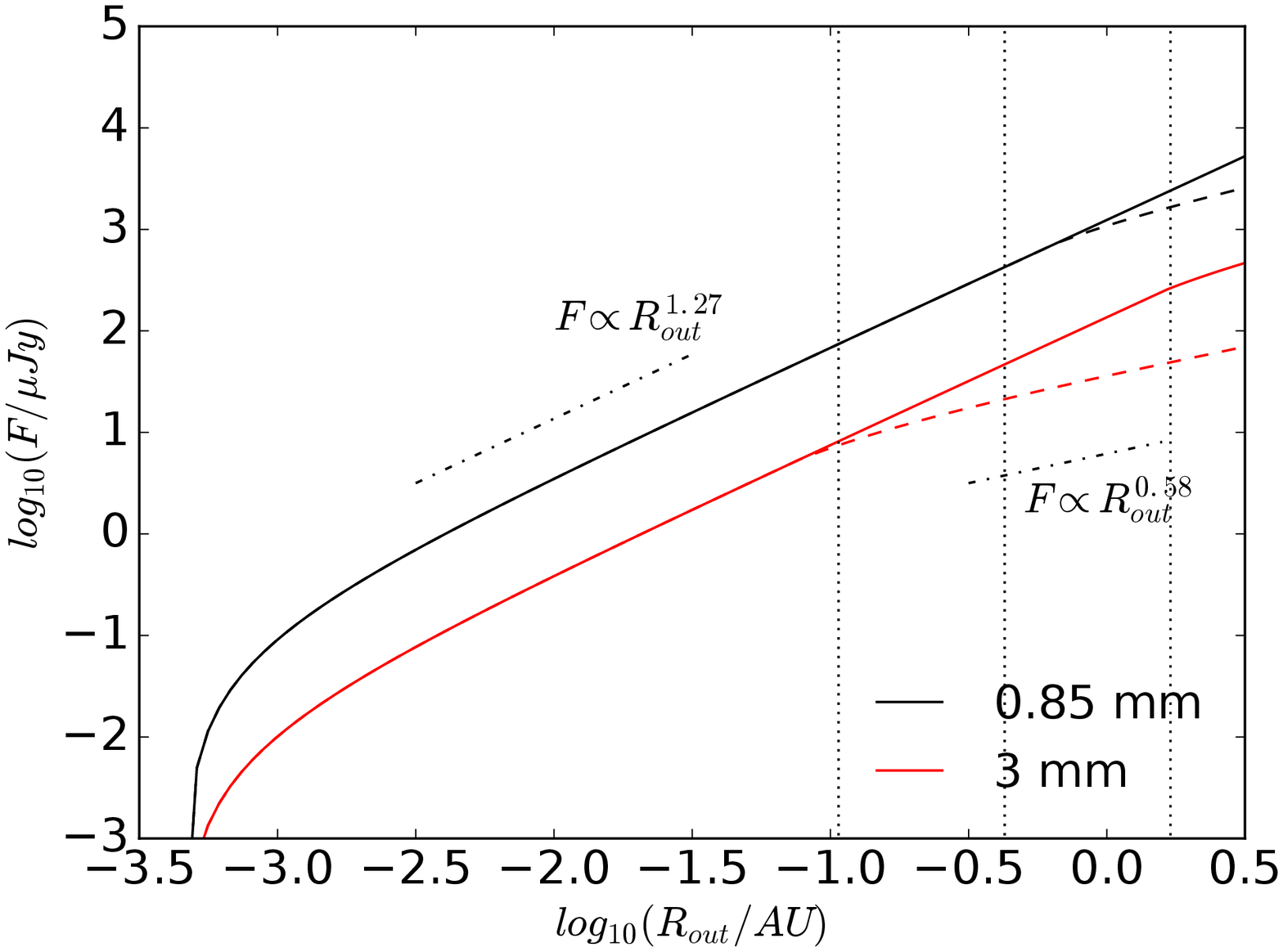} 
\caption{ The integrated flux  as a function of the disc outer radius $R_{out}$. 
$\dot{M_{p}}$ is $10^{-5}M_{J}/yr$ and $M_{p}=1 M_{J}$. The disc is assumed to be viscous heating dominated, and thus 
the planet irradiation is not considered. The black and red curves show the flux
at 0.85 mm and 3mm. The solid curves are for the case with $\alpha=0.001$ while
the dashed curves are for the case with $\alpha=0.01$. The vertical
dotted lines label the three different disc sizes $R_{out}=$0.12, 0.46, and 1.85 AU. 
When the disc becomes optically
thin at a particular wavelength, the flux increases slower with $R_{out}$. The flux is calculated
assuming the CPD is 140 pc away from us. }
\vspace{-0.1 cm} \label{fig:onedint}
\end{figure}

With the effective brightness temperature known, we can calculate the flux at each radius and integrate the total flux in the disc.
Figure \ref{fig:onedint} shows the integrated flux as a function of the disc outer radius $R_{out}$. The solid
curves are for the case shown in the upper panels of Figure \ref{fig:oned}  having a lower $\alpha$ value, 
while the dashed curves are for the case shown in the bottom panels of Figure \ref{fig:oned} having a higher $\alpha$ value. 
The vertical
dotted lines label the three different disc sizes $R_{out}=$0.12, 0.46, and 1.85 AU, which correspond to a CPD around a Jupiter mass planet
at 5, 20, 80 AU from the central solar mass star. 
As clearly demonstrated in Figure \ref{fig:onedint}, when the disc is optically thick at a particular wavelength, the flux is independent
of the disc surface density and the integrated flux roughly scales as $R_{out}^{1.27}$. Thus, a CPD at 80 AU
is almost 34 times brighter than the CPD at 5 AU as long as it is optically thick. On the other hand, when the
disc becomes optically thin, the flux scales as  $R_{out}^{0.58}$. Thus, a CPD at 80 AU will only be 5 times brighter
than the CPD at 5 AU when the disc is optically thin.

\begin{figure} 
\centering
\includegraphics[trim=0cm 0cm 0cm 0cm, width=0.5\textwidth]{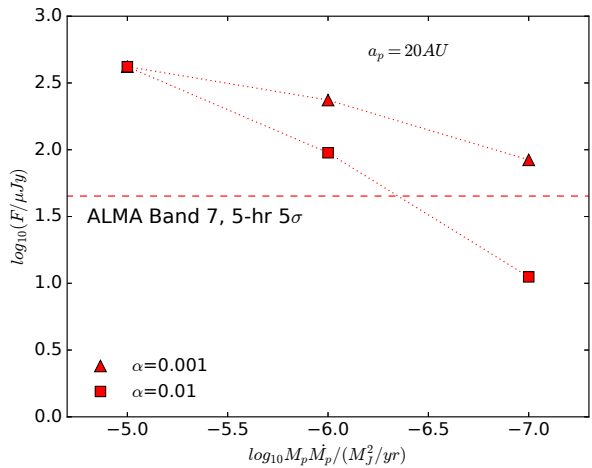} 
\caption{ The flux at 0.85 mm (ALMA band 7) for viscous heating dominated CPDs with different accretion rates.
The systems are assumed
to be 140 pc away from us. Cases
with two different $\alpha$ values are shown.
The planet mass is assumed to be 1 M$_{J}$ and the planet is 20 AU away from the central star ($R_{out}=0.46 AU$).
The horizontal line ( 56 $\mu Jy$) is the 5$\sigma$ detection level  at ALMA band 7 with 5 hours integration.}
\vspace{-0.1 cm} \label{fig:almaone}
\end{figure}

\begin{figure*} 
\centering
\includegraphics[trim=0cm 0cm 0cm 0cm, width=0.9\textwidth]{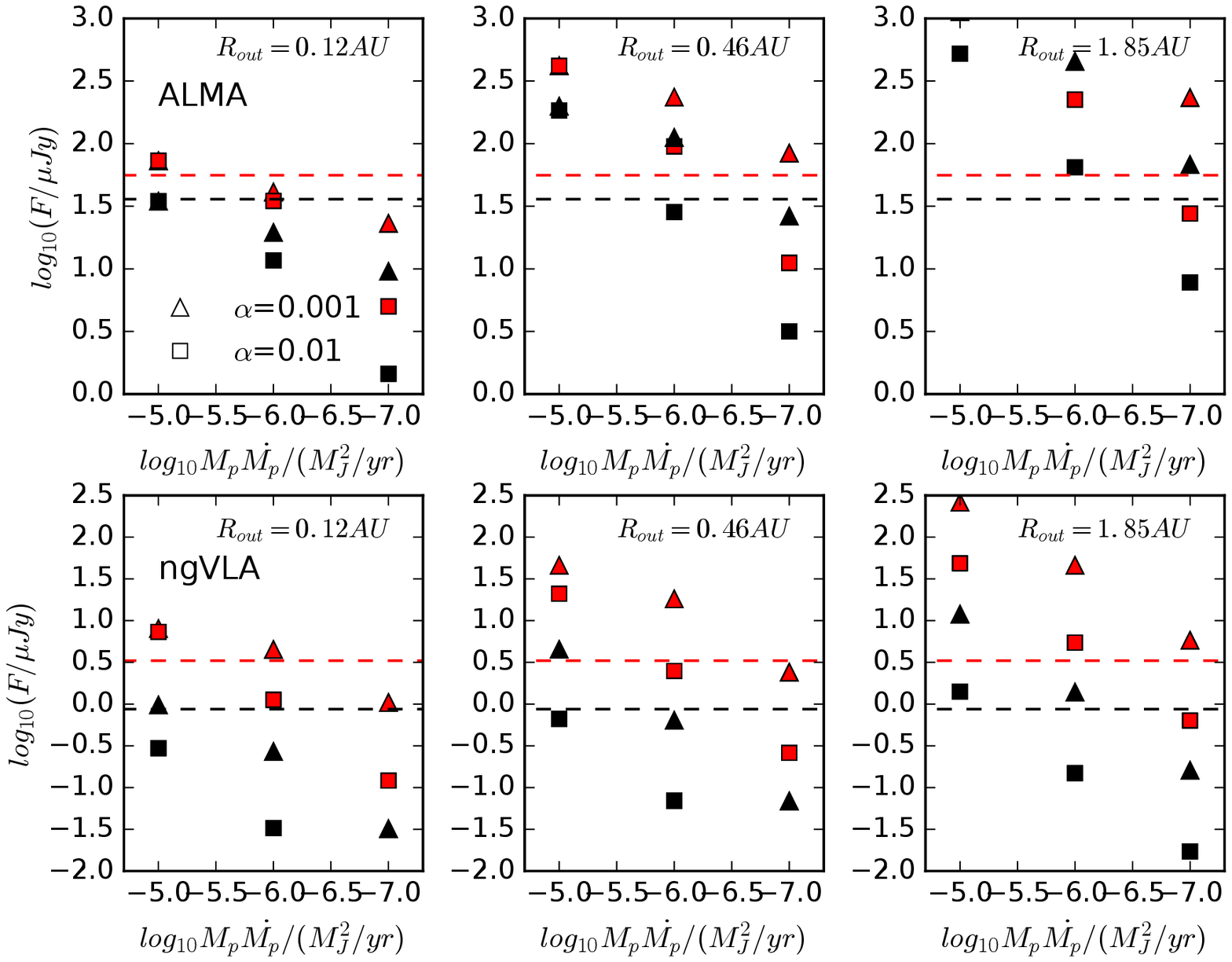} 
\caption{The flux for viscous heating dominated CPDs with different accretion rates.
The upper panels show fluxes at 0.85 mm and 1.3 mm (red and black symbols, respectively). 
The red and black dashed lines indicate the 5$\sigma$ level achieved by ALMA respectively at 0.85 mm 
and 1.3 mm  with 5 hours of integration on source. 
The lower panels show fluxes at 3 mm and 10 mm (red and black symbols, respectively). 
The red and black dashed lines (3.31 $\mu Jy$ and 0.87 $\mu Jy$) 
indicate the 5$\sigma$ level achieved by 5 hours ngVLA observation with the assumed RMS noise level at these two bands 
(1-hour RMS noise of 1.48 $\mu Jy$ at 3 mm and 0.39 $\mu Jy$ at 10 mm).  
From left to right panels, $R_{out}$ are 0.12 AU, 0.46 AU, and 1.85 AU, which correspond
to $a_{p}$ of 5 AU, 20 AU, and 80 AU for a Jupiter mass planet around a solar mass star. Triangles are the cases with $\alpha$=0.001
while squares are the cases with $\alpha$=0.01. }
\vspace{-0.1 cm} 
\label{fig:alma}
\end{figure*}

Figure \ref{fig:almaone} shows the total flux at 0.85 mm (ALMA band 7) for CPDs with $R_{out}=0.48 AU$, 
or equivalently CPDs at $a_{p}=20 AU$ with $M_{p}=1 M_{J}$
and $M_{*}=M_{\odot}$. The flux is scaled by assuming that the systems are 140 pc away from us. 
The two most plausible $\alpha$ values ($\alpha$=0.001, and 0.01 as in Zhu \etal 2016)
are shown. The horizontal line is the 5$\sigma$ detection level (56 $\mu Jy$) achieved by ALMA band 7 observation with 5 hours of integration on source \footnote{Using ESO ALMA sensitivity calculator for a source that transits overhead, the 1-hour RMS noise level of ALMA band 7 (345 GHz) is 25 $\mu$Jy/beam and the 1-hour RMS noise level of ALMA band 6 (230 GHz) is 16 $\mu$ Jy/beam, assuming the maximum bandwidth (7.5 GHz per polarization) for 43 antennas in the ``automatic'' weather conditions. }. 
As  shown in the figure, the flux is independent of the $\alpha$ value when the accretion rate
is high ($M_{p}\dot{M_{p}}=10^{-5}M_{J}^2/yr$), since the discs are optically thick under such high accretion rates. 
When the accretion rate gets lower, the discs become
optically thin at mm/cm and their brightness depends on their surface density.
The discs with lower
$\alpha$ values (triangles) are brighter since they have higher surface density. 
Overall, CPDs at 20 AU should be detectable with ALMA unless the disc's accretion rate is very low
$M_{p}\dot{M_{p}}=10^{-7} M_{J}^2/yr$ and $\alpha$ is large ($\alpha\ge0.01$).

We summarize our results of Table 1 in Figure \ref{fig:alma}. From left to right, $R_{out}$ is 0.12 AU, 0.46 AU, and 1.85 AU, or equivalent
to a CPD around a Jupiter mass planet that is 5 AU, 20 AU, and 80 AU away from a solar mass star. Both Table 1 and Figure  \ref{fig:alma}
clearly show that the disc's radio emission decreases dramatically at longer wavelengths. In the optically thin limit with the above opacity law, the flux will decrease as $\lambda^3$. 
At VLA bands, the disc is normally 1-3 orders of magnitude fainter compared with the flux at ALMA bands. On the other hand, the RMS noise of VLA is less than a factor of 3 smaller than
the RMS noise of ALMA \footnote{Using the VLA exposure calculator, we derive that the 1-hour RMS noise of  the A array Q band observation is  10 $\mu$Jy. }. Thus, ALMA will be much more sensitive to CPDs than VLA. On the other hand, ngVLA may be as sensitive as ALMA for detecting CPDs. 
The upper panels of Figure \ref{fig:alma} show the flux at ALMA
band 7 (red) and 6 (black), while the lower panels show the flux at ngVLA bands at 3 mm and 10 mm. 
The horizontal lines are the 5$\sigma$ detection levels achieved by 5 hours integration at each band. For ngVLA, we assume the 1-hour noise RMS 
is 1.48 $\mu Jy$ at 3 mm and 0.39 $\mu Jy$ at 10 mm.  
Only discs with $\alpha$=0.001 (triangles) and $\alpha$=0.01 (squares)
are shown in the figure. Figure \ref{fig:alma} suggests that it is easier to detect CPDs at shorter radio wavelengths, and
CPDs will also be bigger and brighter when they are further away from the star. With $\alpha=0.001$, $M_{p}=1 M_{J}$, and $M_{*}=M_{\odot}$, 5 hours observation 
can safely detect CPDs that are more than 20 AU away from the central star
and accreting at $\dot{M_{p}}\gtrsim 10^{-7}M_{J}/yr$. If $\alpha=0.01$, 5 hours observation can detect CPDs at 20 AU away from the star and 
accreting at $\dot{M_{p}}\gtrsim 10^{-6}M_{J}/yr$.

\subsection{Irradiated Discs}
\begin{figure*} 
\centering
\includegraphics[trim=0cm 0cm 0cm 0cm, width=0.9\textwidth]{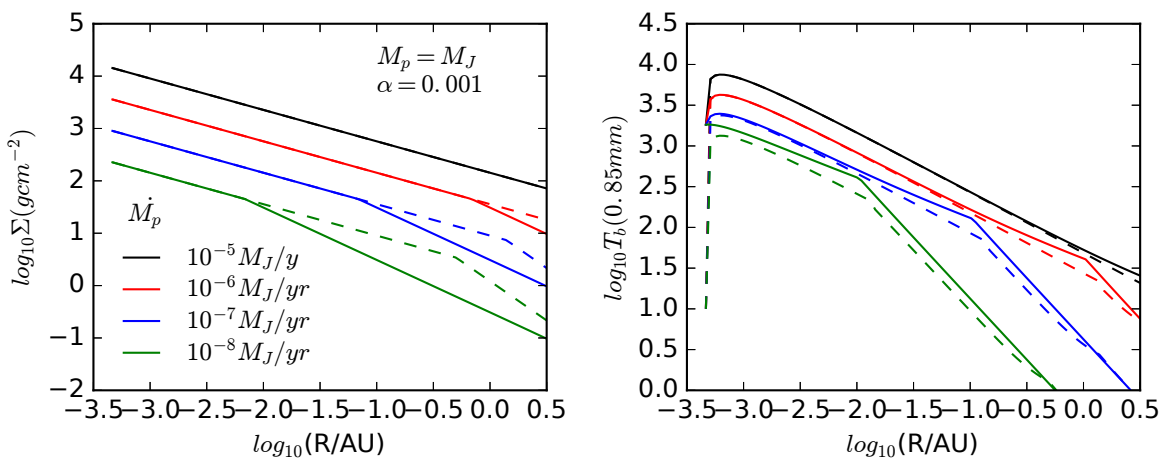} 
\vspace{-5.5 cm} 
\caption{ The radial profiles of the disc surface density and effective brightness temperature for discs without irradiation (dashed curves)
and with irradiation (solid curves). For the irradiated cases, the central planet is assumed to have the luminosity of 0.001 $L_{\odot}$.   }
\label{fig:onedirr}
\end{figure*}

When the irradiation from the planet dominates the viscous heating, $T_{c}$ follows $R^{-1/2}$ and $\Sigma$ follows $R^{-1}$ (Equations \ref{eq:tirr} and \ref{eq:Sigma2}). For comparison, $T_{c}$
and $\Sigma$ of the viscously heated disc follow $R^{-9/10}$ and $R^{-3/5}$ (Equations \ref{eq:Tc} and \ref{eq:Sigma}). The disc structure for both viscous heating dominated and irradiation dominated discs
is shown in the left panel of Figure \ref{fig:onedirr}.  The disc surface density drops more sharply when the irradiation starts dominating the viscous heating. 
On the other hand, the right panel of Figure \ref{fig:onedirr}  shows that the effective brightness temperature and the mm/cm flux do not 
change much between the irradiated and viscous cases. 
This is because $T_{c}\times\Sigma$ follows $R^{-1.5}$ for both the viscously heated and irradiated discs, and 
$T_{c}\times\Sigma$ is proportional to
the effective brightness temperature when the disc is optically thin. 
This looks like an coincident, but it is the direct cause from the viscous accretion theory.
Equation \ref{eq:nusigma} suggests that 
\begin{equation}
T_{c}\Sigma = \frac{\dot{M}\Omega\mu}{3\pi\alpha\Re} \label{eq:tcsigma}
\end{equation}
when $R\gg R_{in}$.
Thus, whatever the heating mechanism is, $T_{c}\times\Sigma$ is unchanged and 
directly determined by $\dot{M}$. As shown in Table 2, the flux difference
between cases with $L_{p}=10^{-3} L_{\odot}$ and cases with $L_{p}=10^{-5} L_{\odot}$ is normally $<10\%$.
In rare cases, the difference can go up to $30\%$.

\section{Discussion}

\subsection{CPD Models With Different Assumptions}
\begin{figure} 
\centering
\includegraphics[trim=0cm 0cm 0cm 0cm, width=0.5\textwidth]{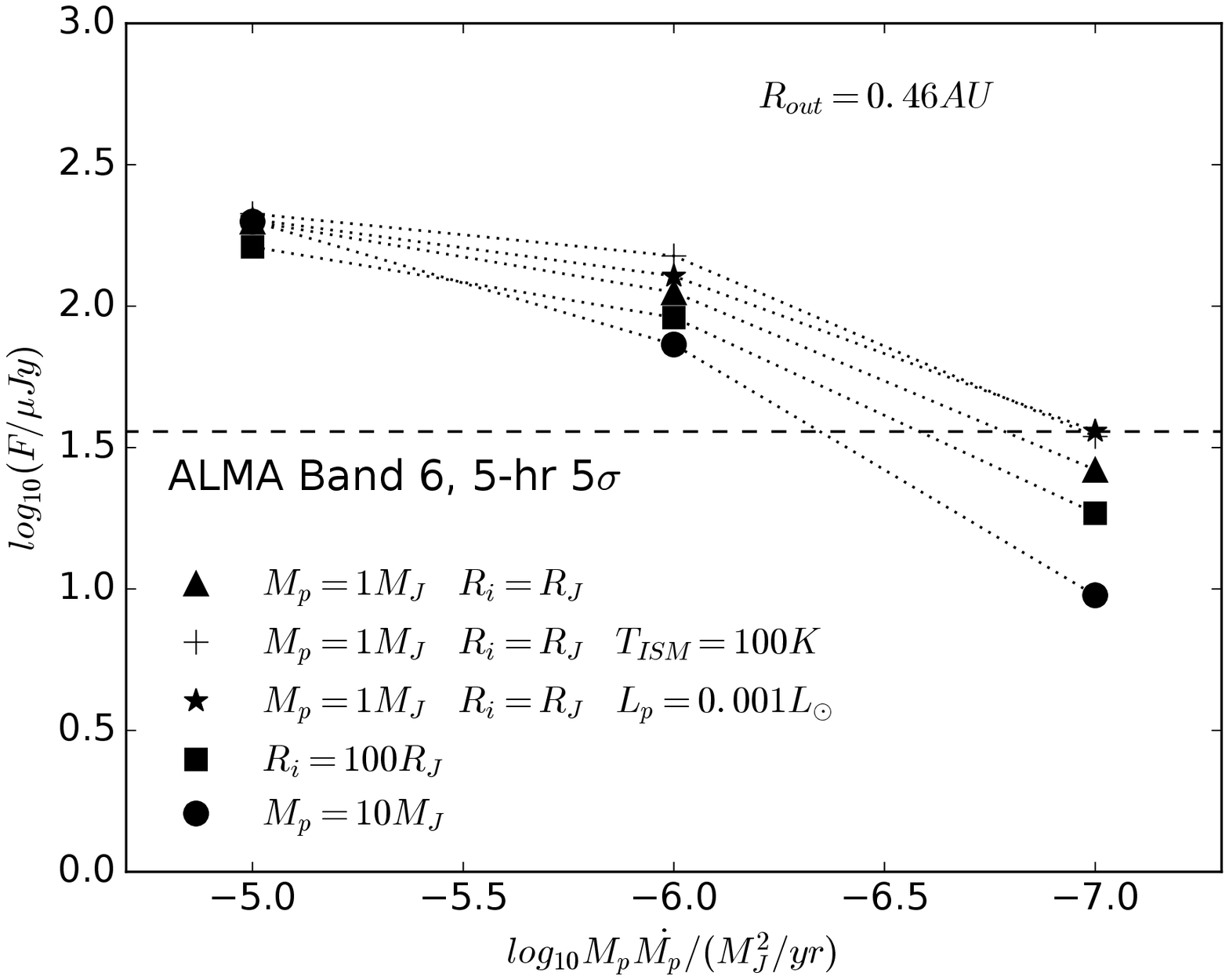} 
\caption{The flux at 1.3 mm (ALMA band 6) for CPDs with different accretion rates.
All these cases have $M_{p}\dot{M_{p}}=10^{-5}M_{J}^2/yr$, $\alpha$=0.001, and $R_{out}=0.46$ AU. 
 Cases with  different background irradiation ($T_{ISM}$), planet irradiation, planet masses ($M_{p}$), and  inner radii ($R_{i}$) are shown.
The horizontal line(36 $\mu Jy$)  is the 5$\sigma$ detection level at ALMA band 6 with 5 hours integration.  }
\vspace{-0.1 cm} \label{fig:variousmodel}
\end{figure}

Our calculations suggest that CPDs should be detectable by ALMA (note the caveat in \S 4.6). 
Here we will explore how the calculated flux will change 
if we relax some of our assumptions. 

Our disc model in \S 2 (Equation \ref{eq:Fv}) assumes a zero torque
inner boundary, which means that the disc rotates at the local Keplerian speed at the inner boundary.
In reality, the Keplerian rotating disc joins the slowly rotating planet through a boundary layer
or  magnetosphere channels. Unless the planet has strong magnetic fields (more discussion in
Zhu 2015), a boundary layer will form between the disc and the planet, where half of the accretion energy
($L_{acc} =GM_{p}\dot{M_{p}}/R_{p}$) is released. This boundary layer could irradiate and heat up the CPD.
To include this effect, we set $L_{irr}=0.5 L_{acc}$ in Equations \ref{eq:Fv}-\ref{eq:Sigma2}. 
The resulting mm/cm flux from such discs
are shown on the right side of each $T_{b}$ and flux column in Table 1. We can see that the irradiation has
little effect on the mm emission. Only when the disc accretion rate is low and $\alpha$
is large (in other words, the surface density is low), the irradiation can increase the total flux by 20\%. 

If the CPD is close to the star, 
the irradiation from the star and the circumstellar disk can be significantly stronger. At 5 AU, the circumstellar disc has a temperature of 
$\sim$ 100 K. We thus calculate the radio emission from CPDs with $T_{ISM}$=100 K which mimics the irradiation from
the star and the circumstellar disc. The fluxes are shown as the plus signs in Figure \ref{fig:variousmodel}. As expected from the argument in \S 3.2,
such strong irradiation has little effect on the disc's radio emission.  

The irradiation onto the disc can also be strong if the planet is bright by itself (e.g. it has a ``hot start'').
But as shown in Table 2 and \S 3.2, irradiation does not change the mm emission significantly 
because the flux in the optically thin limit only depends on the accretion rate and is independent of the heating mechanism. 
The mm emission from the CPDs irradiated by a $L=0.001 L_{\odot}$ planet is shown in Figure \ref{fig:variousmodel} as the star signs
and they are quite similar to the non-irradiated case.

Then, we explore how  the mm emission changes if we vary $R_{in}$ in the model. 
The size of a young Jupiter mass planet could be larger than $R_{J}$ if it  starts with a
high internal energy \citep{Marley2007}.  A young planet may also have strong magnetic fields, truncating the CPD
far away from the planet. For the emission at near-IR and mid-IR, Zhu (2015) has shown that the IR flux
is very sensitive
to $R_{in}$ since most near-IR/mid-IR emission comes from the inner disc. On the other hand, 
we expect CPD's mm/cm flux is sensitive
to $R_{out}$ instead of $R_{in}$, since most mm/cm emission comes from the outer disc. 
We demonstrate this point in Figure \ref{fig:variousmodel}. Even with $R_{i}$=100$R_{J}$,
the mm/cm flux is almost unchanged. 

Our models assume that $R_{out}$ is 1/3 of the Hill radius, which is supported by many theoretical studies \citep{QuillenTrilling1998,MartinLubow2011}. 
However, in case $R_{out}$ is smaller than this value by a factor of 2, the total flux will only decrease by a factor of 2.4 if the
disc is optically thick or by a factor of 1.5 if the disc is optically thin (the scaling law is given in Figure \ref{fig:onedint}).

Finally, we study how the mm/cm flux changes with different $M_{p}$. Since the disc's effective
temperature only depends on $M_{p}\dot{M_{p}}$, at a given $M_{p}\dot{M_{p}}$, the  flux will be independent of $M_{p}$
 when the disc is optically thick at mm/cm. On the other hand,
$T_{c}\Sigma$ depends on $\dot{M_{p}}M_{p}^{1/2}$ (Equation \ref{eq:tcsigma}). Thus,
when the disc is optically thin, mm/cm flux should roughly scale with $M_{p}^{-1/2}$ at a given $M_{p}\dot{M_{p}}$. 
This has been shown in Figure \ref{fig:variousmodel}.  With $M_{p}\dot{M_{p}}=10^{-5}M_{J}^2/yr$ fixed,
the flux from the CPD around a $10 M_{J}$ planet is almost the same as the flux from the CPD around a 1 $M_{J}$ planet since the disk is optically thick.
But with $M_{p}\dot{M_{p}}=10^{-7}M_{J}^2/yr$, the flux from the $M_{p}=10M_{J}$ case is a factor of
$\sim$3 smaller than the $M_{p}=1 M_{J}$ case.

Overall, our calculated mm/cm flux is robust and it is not affected by the boundary layer, the irradiation, the inner disc radius, and only mildly
affected by $M_{p}$.

\subsection{Radio Flux From Minimum Mass Sub-nebula Model}
\begin{figure*} 
\centering
\includegraphics[trim=0cm 0cm 0cm 0cm, width=0.9\textwidth]{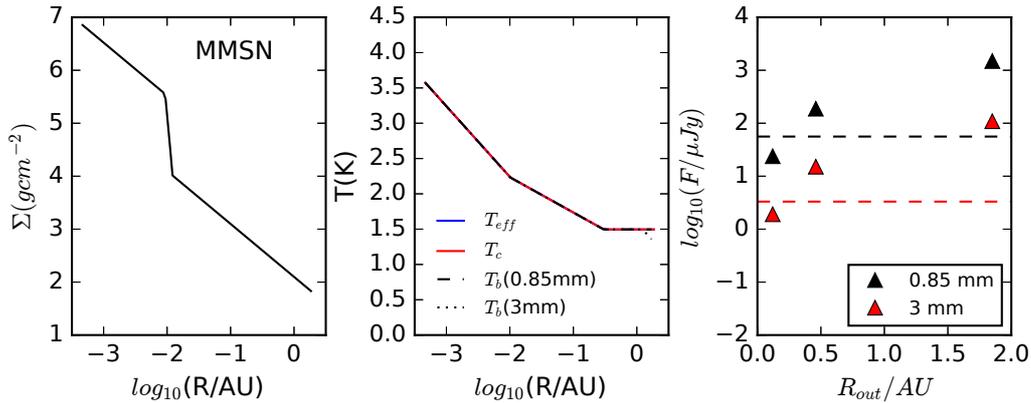} 
\vspace{-5.5 cm} 
\caption{The minimum mass sub-nebula model and its mm/cm emission. The disc surface density and 
temperature \citep{MosqueiraEstrada2003} are shown in the left and middle panel. The middle panel
also shows the brightness
temperature at 0.85 mm and 3 mm. Since the disc has a uniform temperature vertically and the disc
is optically thick, the effective brightness temperature equals the disc midplane temperature.  The integrated mm/cm emission
for discs with different $R_{out}$
are shown in the right panel. The horizontal lines are the 5-hr 5$\sigma$ detection levels for ALMA and ngVLA. }
\label{fig:mmsn}
\end{figure*}

Although we have little observational constraints on CPDs around young planets,
we can use satellites around Jovian planets in our solar system to study CPDs that were in
our solar system. Several CPD models  have been constructed to understand satellite formation in our solar system 5 billion years ago. 
In one model, by spreading the mass of Galilean satellites into a mini disc around Jupiter, 
the ``minimum mass sub-nebula''  disc was constructed \citep{LunineStevenson1982, MosqueiraEstrada2003}.
Such model has a high gas surface 
density of few$\times 10^{5}$ g cm$^{-2}$. The disc surface density and temperature in \cite{MosqueiraEstrada2003} 
are plotted in the left and middle panel of Figure \ref{fig:mmsn}. In this model, 
the disc is assumed to have the same temperature along the vertical direction. 
We calculate the disc effective brightness temperature
 and integrate the total mm/cm emission from the disc 
(the right panel of Figure \ref{fig:mmsn}). Surprisingly, such a disc at 5 AU 
(so $R_{out}\sim$0.12 AU) is even detectable by ALMA. Thus, if the ``minimum mass sub-nebula''  disc 
once existed around our Jupiter, we will be able to detect such a disc in a young solar 
system analog 140 pc away. In reality, the circumstellar material may confuse the detection even with ALMA's  highest resolution unless
a deep gap has been carved out (more discussion in \S 4.5).
 On the other hand, the other satellite formation model, 
so-called ``gas-starved'' disc model \citep{CanupWard2002}, has a much lower surface density. 
The ``gas-starved'' disc model is 
basically an $\alpha$
disc model with $\dot{M_{p}}\sim2\times10^{-7}M_{J}/yr$, so its mm/cm emission can be found
from our main set of models. We will be able to detect such CPDs if they are 20 AU away from
the central star with $\alpha<0.01$.

\subsection{Radio Emission from CPD Jets and Winds}
Most astrophysical accretion discs, from AGNs to protostars, 
produce jets and winds that are observable at both radio and optical wavelengths. 
Thus, accreting CPDs could also produce jets and winds. MHD simulations by \cite{Gressel2013}
have indeed suggested that disc winds can be launched in CPDs. 
To estimate the centimetre free-free emission of jets launched in CPDs, we extrapolate the relationship for 
protostellar jets given in \cite{Anglada2015},
\begin{equation}
\frac{F_{\nu}d^2}{mJy\,\, kpc^2}=0.008\left(\frac{L_{bol}}{L_{\odot}}\right)^{0.6}\,\,\label{eq:fnujet}
\end{equation}
to the planetary mass regime. 
With $L_{bol}=1.57\times 10^{-3} L_{\odot}$ in our $M_{p}\dot{M_{p}}=10^{-5} M_{J}^2/yr$ model,
we can estimate that $F_{\nu}$ from the jet is 8.48 $\mu Jy$ at cm  if d=140 pc. The flux of 8.48 $\mu Jy$ is undetectable
by ALMA, but it can be detected by ngVLA. Thus, when we have CPD detections with ngVLA in future,
we may need to separate the flux from the disc
and the flux from the jet using the radio spectral index, similar to the practice we are carrying out 
in the circumstellar discs (e.g. \citealt{CarrascoGonzalez2016}). 

We caution that such extrapolation using Equation \ref{eq:fnujet} is extremely crude without any 
physical argument to support. The extrapolated flux can be off by more than one order of magnitude. 
Nevertheless, this provides a rough estimate of the possible radio emission from the disc. 

\subsection{Constraining CPD properties}
\begin{figure*} 
\centering
\includegraphics[trim=0cm 0cm 0cm 0cm, width=0.47\textwidth]{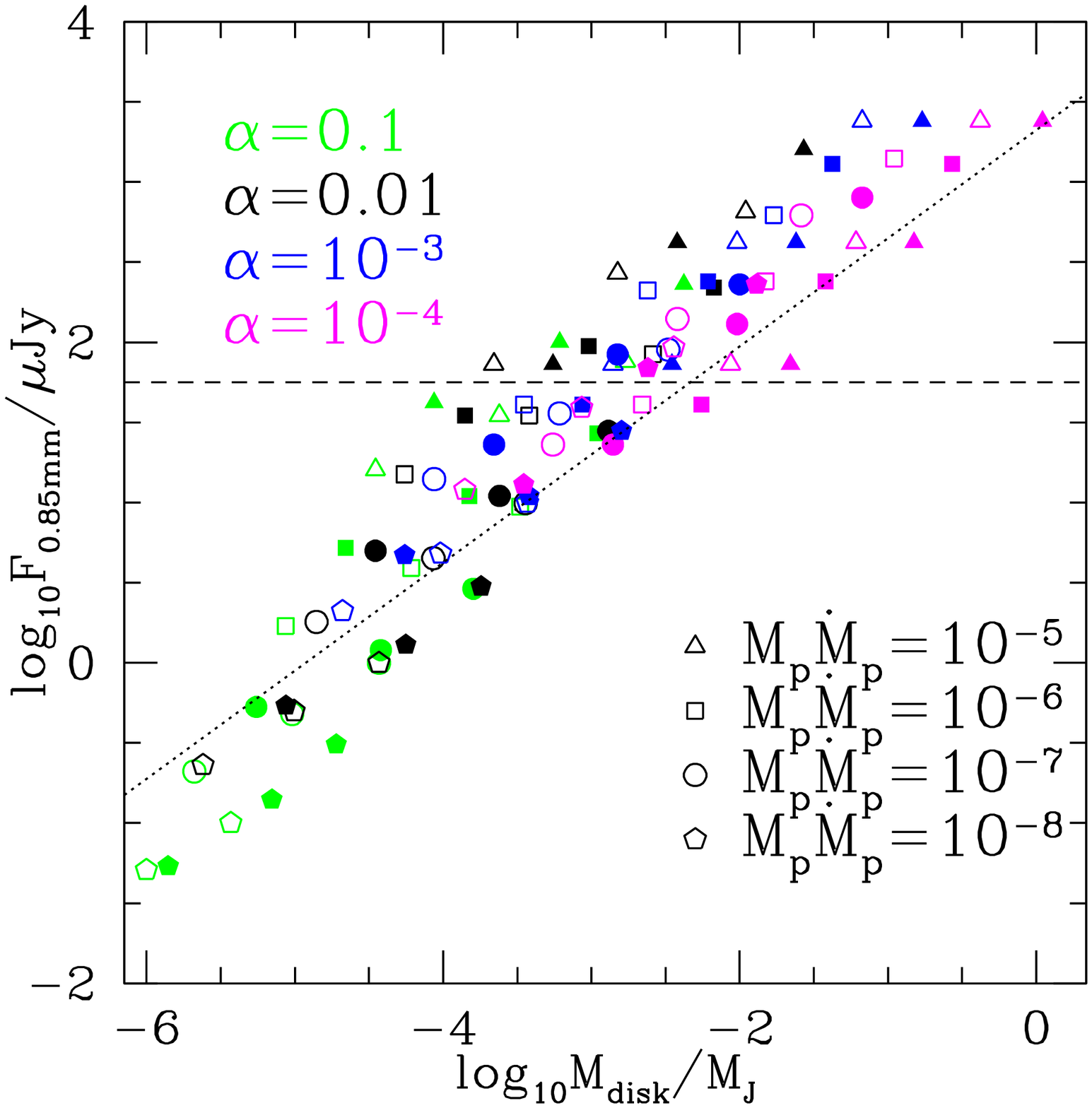} \
\includegraphics[trim=0cm 0cm 0cm 0cm, width=0.51\textwidth]{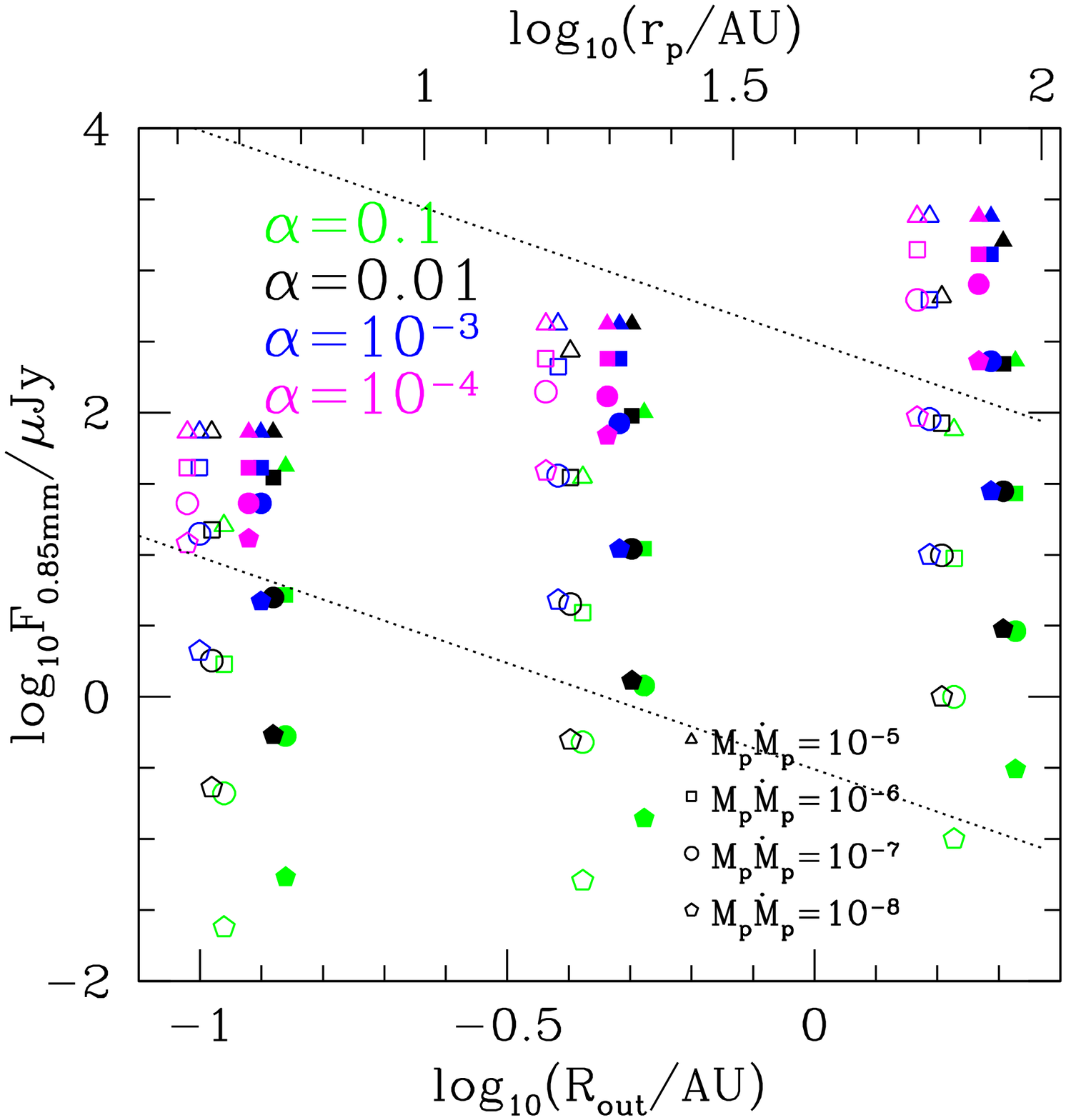} 
\caption{ Left: The disc mass and submm emission (at 0.85 mm) for our models. 
Right: The disc outer radius and submm emission (at 0.85 mm) for our models. 
The filled points are
discs with $M_{p}=1 M_{J}$ (from Table 1 without boundary irradiation) while the open points are discs with $M_{p}=10 M_{J}$
(from Table 2 with little irradiation, $L_{p}=10^{-5} L_{\odot}$).  In the right panel, the distance of the planet from the central star is plotted
on the top axis (assuming a Jupiter mass planet around a solar mass star) and the three $R_{out}$ (0.12, 0.46, 1.85 AU) are slightly shifted horizontally to show overlapping points. 
The horizontal line in the left panel is the 5-hr 5$\sigma$ detection level for ALMA Band 7. The two dotted lines in the right panel represent the flux from the fiducial circumstellar disc (Equation \ref{eq:circumstellar}) and
the flux which is a factor of 1000 smaller than the fiducial case, assuming the ALMA Band 7 observation with the 0.03'' resolution. 
}
\label{fig:mass}
\end{figure*}

If there is ALMA detection of discs around planetary mass objects,
we can use our disc models to constrain CPD properties.
We  take the recent ALMA detection of the disc around OTS44
(a 12$M_{J}$ object) as an example \citep{Bayo2017}. OTS44 accretes at 8$\times10^{-12}M_{\odot}/yr$.
Thus, $M\dot{M}\sim10^{-7}M_{J}^2/yr$. The source is detected at $\sim 100\mu$Jy at ALMA Band 6. 
Since the disc is unresolved by ALMA with the beam of 0.16", the disc size is smaller than
13 AU with the distance of 160 pc. All our models satisfy this size constraint. 
By using Table 2  and under the $M_{p}=10 M_{J}$,  $M\dot{M}\sim10^{-7}M_{J}^2/yr$, $L=0.001 L_{\odot}$ 
category, we can find that 
the disc mass is around $0.01$ $M_{J}$ and $R_{out}\sim 1$AU if $\alpha=10^{-4}$.  If $\alpha=10^{-3}$, the CPD with $R_{out}=1.85$ AU
has a flux of 28 $\mu$Jy and a mass of 1.3$\times10^{-3}M_{J}$. As discussed in \S 3.1, the flux scales as $R_{out}^{0.58}$ when 
the disc is optically thin, 
and the mass scale as $R_{out}^{1}$ for strongly irradiated discs. Thus, to emit $100 \mu$Jy, 
the disc has a outer radius of 17 AU and also a mass of  $0.01 M_{J}$ with $\alpha=10^{-3}$.

The disc mass and mm emission for all our models are given in the left panel of Figure \ref{fig:mass}. 
We can see that the 5 hours ALMA Band 7 observation can probe CPDs having 10$^{-4}$ $M_{J}$ of gas, which is equivalent to
0.026 lunar mass of dust assuming the dust-to-gas mass ratio of 1/100.
There is a loose correlation between the disc mass and the mm emission. Thus, by using such
correlation, we can estimate the disc mass from its mm emission. 
From Table 2, we see that the flux at ALMA Band 7 is roughly 3 times larger than the flux at Band 6.
Thus, for OTS 44, the flux at ALMA Band 7 should be around 300 $\mu$Jy.
Using Figure \ref{fig:mass}, the estimated disc mass for OTS 44 is again $0.01 M_{J}$.

On the other hand, OTS 44 is a free floating planetary mass object whose outer disc radius is unconstrained. 
For a planetary mass object orbiting around a central star,  its outer radius is limited to 1/3 of its Hill radius. Thus, knowing
the planet's position with respect to the central star, $R_{out}$ can also be estimated with some assumptions about the planet and the stellar mass. 
Then, both
the CPD's radio flux and $R_{out}$ are known.  We plot the mm emission and $R_{out}$ for all our models in the right panel of
Figure \ref{fig:mass}. 

We can apply this figure to the recent ALMA tentative detection of the point source within HD 142527 \citep{Boehler2017}.
HD 142527 system is $\sim$156 pc away from us.
The point source candidate  is at 50 AU from the central star and has the flux of 0.80 mJy at 0.88 mm . 
The right panel of Figure \ref{fig:mass} suggests that the upper flux limit from the CPDs at $\sim$ 50 AU is $\sim$ 2 mJ. Thus,
 the detected flux is consistent with the emission from the CPD. Since the measured flux is close to the upper limit of all the models, 
the CPD candidate is almost optically thick at mm and the $M_{p}\dot{M_{p}}$ should be $\gtrsim10^{-7}M_{J}^2/yr$ based on the figure. If the inner radius of the CPD
is 1 $R_{J}$, the CPD should be brighter than 17.2 magnitude at the mid-IR M band  \citep{Zhu2015}, which can be studied by JWST in future. 

We can also use the right panel of Figure \ref{fig:mass} to constrain the
non-detection of discs around GSC 0614-210 b \citep{Bowler2015},
DH Tau b \citep{Wolff2017} and GQ Lup B \citep{MacGregor2017, Wu2017} by ALMA. The accretion rates
of these discs have been measured with various techniques. 
 DH Tau b has an accretion rate of 3.2$\times$10$^{-12}M_{\odot}/yr$ \citep{Zhou2014}, 
GQ Lup B has an accretion rate of $10^{-12}$ to $10^{-11} M_{\odot}/yr$ \citep{Wu2017} (but see \citealt{Zhou2014} which derive an accretion
rate of $10^{-9.3}M_{\odot}/yr$.), and  GSC 0614-210 b has an accretion rate of $10^{-10.8}M_{\odot}/yr$ \citep{Zhou2014}.
Thus, $M\dot{M}$ in these discs are $\sim10^{-8}-10^{-7}M_{J}^2/yr$. Assuming these discs fill 1/3 of their Hill radius,
the non-detection may suggest that $\alpha\gtrsim0.001$ in these discs.

HD 169142b is another CPD candidate 20 AU away from the central 1.65 $M_{\odot}$ star. If its L band emission comes from the accreting CPD \citep{Reggiani2014},
\cite{Zhu2015} estimate that $M_{p}\dot{M_{p}}$ is $10^{-5}M_{J}^2$/yr.  Then Table 1 suggests that the disc should be around 400 $\mu$Jy at ALMA Band 7, which
is consistent with the estimate from numerical simulations in \cite{Zhu2016}.
Finally, we caution that
we have assumed perfect coupling between dust and gas so far. If we relax this assumption, the discs without radio detections  may still have
a lot of gas despite very little dust, and HD 169142b may be faint at radio bands (more discussion in \S 4.6).

\subsection{The Resolution Requirement}

In most above cases, the planet/brown dwarf is not embedded in the circumstellar disc so 
there is little confusion between the tiny disc around the brown dwarf/planet and the circumstellar disc around the central star.
In reality, we want to discover CPDs within circumstellar discs or gaps of circumstellar  discs. Then, we need to not only  
detect CPDs but also have enough spatial resolution to separate them from circumstellar discs. The emission from the background
circumstellar disc or cavity/gap can be simply estimated assuming the disc is optically thin, which is normally a good approximation
for the circumstellar disc region beyond 10 AU. Suppose the circumstellar disc has a surface density and temperature structure of $\Sigma(r)$ and $T(r)$,
the effective brightness temperature will be $t_{b}(r)=T(r)\times\kappa_{mm}\times\Sigma(r)$. If we choose a fiducial model with $\Sigma(r)=$1.78 (r/100 AU)$^{-1}$g cm$^{-2}$ and 
$T(r)=22.1 (r/100 AU)^{-1/2}$ K, which is a $\alpha=0.01$ disc around a solar mass star accreting at $10^{-8}\msunyr$ and roughly consistent with the protoplanetary discs
in Ophiuchus \citep{Andrews2009}, we have
\begin{equation}
t_{b}(\lambda)=1.34\times \frac{\Sigma(r)}{1.78 g cm^{-2}}\frac{T}{22.1 K}\frac{0.87 mm}{\lambda}\,.
\end{equation}
Then, assuming the mm beam size is $\theta$ and the source is 140 pc away, the radio flux will be
\begin{equation}
F(\lambda)=87 \mu Jy\times  \frac{\Sigma(r)}{1.78 g cm^{-2}}\frac{T}{22.1 K}\frac{0.87 mm}{\lambda}\left(\frac{\theta}{0.03''}\right)^{2}\,.\label{eq:circumstellar}
\end{equation}
With $\Sigma(r)=$1.78 (r/100 AU)$^{-1}$g cm$^{-2}$ and 
$T(r)=22.1 (r/100 AU)^{-1/2}$ K, $F(\lambda)$ follows $r^{-1.5}$ and it is plotted in the right panel of Figure 8 assuming the 0.03'' resolution. Clearly
with such high disc surface density, it is difficult to detect any CPD within 20 AU with 0.03'' resolution.  
To detect CPDs at 20 AU, the radio flux from the circumstellar disc within the telescope beam needs to be suppressed by at least 2-3 orders of magnitude.
Such suppression can be physical due to the decrease of the surface density from the gap opening process, or can be observational such as using
a smaller telescope beam. However, there is a limit of using a smaller beam. When the CPD starts to be spatially resolved, decreasing the beam size won't increase
the contrast between the CPD and the circumstellar discs. For the CPD at 20 AU away from the central star, $R_{out}$ is 0.46 AU so that this limiting
spatial resolution is 0.46 AU/140 pc = 0.0033''.  We will achieve the maximum contrast between the CPD and the background circumstellar disc at this resolution. 
This is where ngVLA has more advantage than ALMA with similar sensitivity level but much higher angular resolution. 

Another advantage of using a high spatial resolution is that we can separate the gap region from the smooth circumstellar disc. A giant planet at $r_{p}$ can induced a gap
which can extend from 0.5 $r_{p}$ to 2 $r_{p}$. If we can spatially resolve the gap, we can gain the maximum contrast between the gap and the CPD.
We can also gain a higher contrast if the CPD is at the outer disc where the circumstellar disc has a lower surface density and a lower temperature, as shown by the dotted line
in the right panel of Figure \ref{fig:mass}). 
Thus, to detect CPDs, we should observe discs with large gaps/cavities using the highest resolution possible.  

\subsection{The drift of mm/cm dust particles}
So far we have assumed that dust and gas are perfectly coupled. In reality, dust particles
feel the aerodynamic drag force from the gas and drift radially in discs. The aerodynamic drag force
depends on particle size and  is normally parameterized using the  dimensionless dust stopping time ($T_{s}$).
The radial drift speed is
\begin{equation}
v_{R,d}=\frac{T_{s}^{-1}v_{R,g}-\eta v_{K}}{T_{s}+T_{s}^{-1}}\,, \label{eq:eqvrd}
\end{equation}
where $v_{R,g}$ is the gas radial velocity, $v_{K}$ is the midplane Keplerian velocity, and 
$\eta=-(R \Omega_{K}^{2}\rho_{g})^{-1}\partial P_{g}/\partial R$ is the ratio between the gas pressure gradient
and the stellar gravity in the radial direction \citep{Weidenschilling1977, TakeuchiLin2002}.

When the particle size is smaller than the mean free path of a molecule (which is the case for the problem 
studied here), the drag force is in the  Epstein regime and $T_{s}$ can be written as (e.g. \citealt{espaillat2014})
\begin{equation}
T_{s}=1.55\times 10^{-3}\frac{\rho_{p}}{1 g cm^{-3}}\frac{s}{1 mm}\frac{100 g cm^{-2}}{\Sigma_{g}}\,,\label{eq:ts}
\end{equation}
where $\rho_{p}$ is the dust particle density, $s$ is the dust size, and $\Sigma_{g}$ is the disc gas surface density.

With Equations \ref{eq:eqvrd} and \ref{eq:ts}, we can calculate the dust drift timescale 
\begin{equation}
\tau_{drift}=\frac{R}{v_{R,d}}=\frac{T_{s}+T_{s}^{-1}}{2\pi \frac{\partial ln P}{\partial ln R}\left(\frac{H}{R}\right)^2}T_{orb}\,,
\label{eq:taudrift}
\end{equation}
where $T_{orb}=2\pi/v_{K}$.
For irradiation dominated discs, we have $\Sigma=\Sigma_{0}(R/AU)^{-1}$ and $T= T_{0}(R/AU)^{-0.5} $.
 For the particular case where $M_{p}=1 M_{J}$,
 $\dot{M_{p}}=10^{-8} M_{J}/yr$, and $\alpha=0.001$ (as in Figure \ref{fig:onedirr}), 
 we have $\Sigma_{0}=0.2 g cm^{-2}$ and $T_{0}=40 K$.
Putting these values into Equation \ref{eq:taudrift} and assuming $T_{s}<1$, we derive the particle drift timescale 
\begin{equation}
\tau_{drift}=47.8 \frac{1 mm}{s}\frac{\Sigma_{0}}{0.2 g cm^{-2}}\frac{40 K}{T_{0}}\left(\frac{M_{p}}{10 M_{J}}\right)^{1/2} yrs \,,
\end{equation}
which is extremely short. 
For a comparison, the gas accretion timescale is
\begin{equation}
\tau_{gas}=\frac{R^2}{\nu}=10^5 \left(\frac{M_{p}}{10 M_{J}}\right)^{1/2}\frac{R}{AU}\frac{0.001}{\alpha}\frac{40 K}{T_{0}} yrs\,.
\end{equation}

\begin{figure*} 
\centering
\includegraphics[trim=0cm 0cm 0cm 0cm, width=0.9\textwidth]{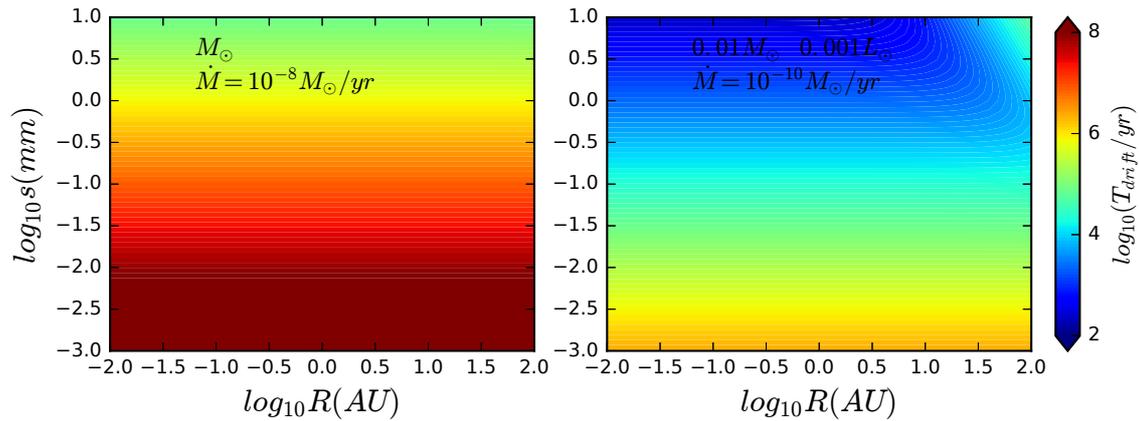} 
\vspace{-5.5 cm} 
\caption{ The drift timescale for the disc around a solar mass object having $L=L_{\odot}$ (the left panel)
and the disc around a 0.01 $M_{\odot}$ object having $L=0.001 L_{\odot}$ (the right panel).  In both cases,
the disc is irradiation dominated and $\alpha$ is assumed to be 0.001. }
\label{fig:drifttime1}
\end{figure*}

The drift timescale of mm particles in CPDs is 3 orders of magnitude smaller than the drift timescale of mm particles in circumstellar disks.
For the circumstellar disc around a solar mass star, the central object  mass is $1000 M_{J}$, $T_{0}$ is $\sim$ 200 K,  $\Sigma_{0}$ is $\sim$1000 g cm$^{-2}$ if $\alpha=0.001$.
Thus, $\tau_{drift}$ is $\sim5\times10^5$ yrs for 1 mm dust particles. The drift timescale for the disc around a solar mass star
 is shown in the left panel of Figure \ref{fig:drifttime1}. 
As the disc's surface density decreases and the mass of the central object  drops,
$\tau_{drift}$ decreases significantly. For the 10 M$_{J}$ planet accreting at a low rate (shown
in the right panel of Figure \ref{fig:drifttime1}), the drift timescale for 1 mm particles is only 1000 years. 
This fast radial drift is consistent with the lack of mm emission from GSC 0614-210 b \citep{Bowler2015},
DH Tau b \citep{Wolff2017} and GQ Lup B \citep{MacGregor2017, Wu2017}. It also suggests that there may still be a lot of gas in these discs, which can be probed by future ALMA molecular line observations. 

\begin{figure*} 
\centering
\includegraphics[trim=0cm 0cm 0cm 0cm, width=0.9\textwidth]{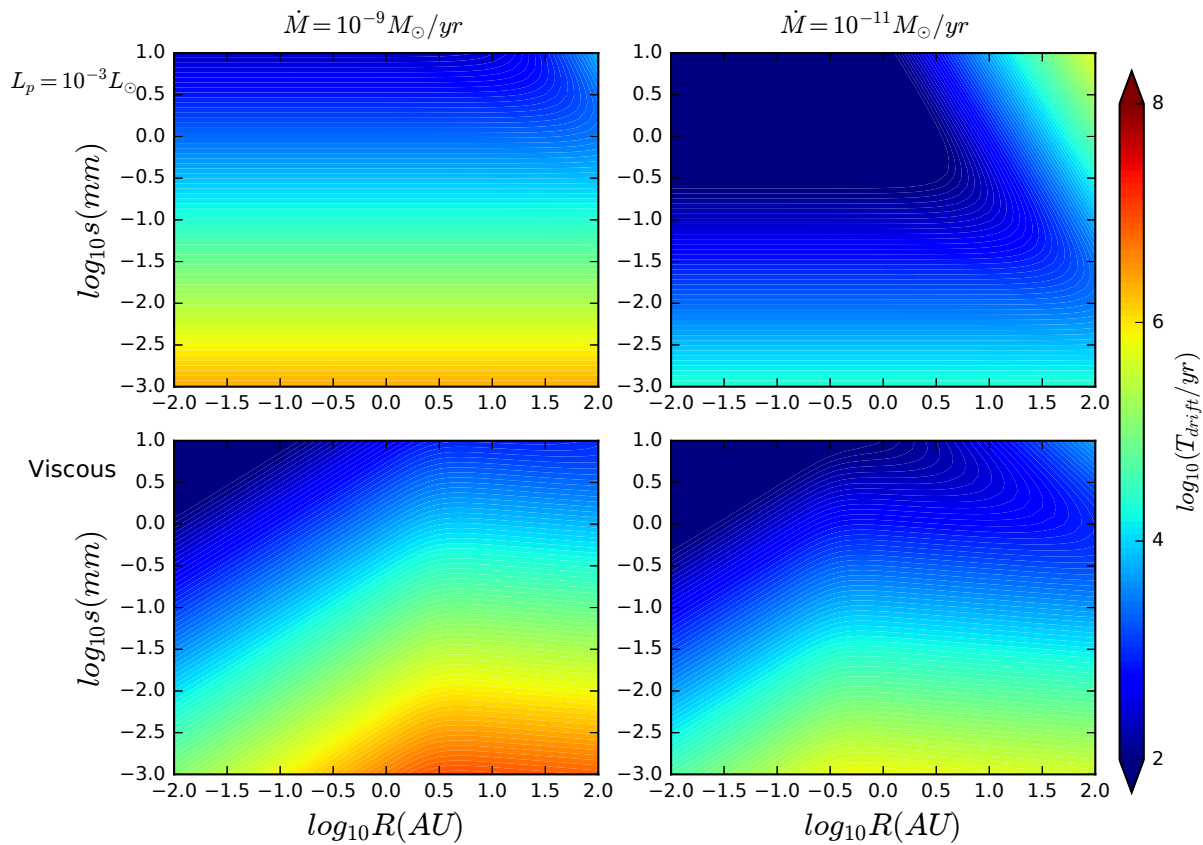} 
\vspace{-0.5 cm} 
\caption{ The drift timescale for discs around a 1 $M_{J}$ planet. The discs
are either irradiation dominated ($L_{irr}=10^{-3} L_{\odot}$, upper panels) 
or viscous heating dominated ($L_{irr}=0$, bottom panels. ). Discs that accrete
at high and low rates are shown in the left and right panels. }
\label{fig:drifttime2}
\end{figure*}

Figure \ref{fig:drifttime2} shows the drift timescale for discs around a  Jupiter mass planet.
The 1 mm particles will drift to the central star within 1000 years
even in discs accreting at $10^{-9}M_{\odot}/yr$. The drift timescale is $<100$ years when the disc accretes
at $10^{-11}M_{\odot}/yr$.  The upper panels show the irradiation dominated cases while the bottom
panels show the viscous heating dominated cases.
Clearly, independent of the heating mechanism, the drift timescale for mm particles
is very short \footnote{When particles settle to the disc midplane or have significant pile up, the feedback from dust to gas becomes important.
Then streaming instability \citep{YoudinGoodman2005} will operate to regulate the dust to gas mass ratio, and dust will still drift fast radially ( slower by a factor of 2 than the speed without including the feedback \citep{BaiStone2010}).}.

The meter barrier in protoplanetary discs becomes the millimetre barrier in circumplanetary discs.

Thus, it is unlikely that we will detect mm/cm signals from discs around planetary mass objects unless:

1) There is a significant population of micron sized dust. At mm/cm wavelengths, the opacity of micron-sized dust is only
several times smaller than that of mm-sized dust \citep{DAlessio2001}. 
Thus, if all the dust in the CPD is in the micron-sized range, 
the CPD can still have detectable radio emission, although the
flux will be weaker than our current estimate. 

 2) Or the disc has a high surface density so that the mm-sized dust has a smaller stopping time and the drift timescale is long.
This means that the disc either
accretes at a high rate or the $\alpha$ is small. 

3) Or dust is continuously being replenished into the disc. The replenishment could come from the accretion
streams from the circumstellar disc, or infalling envelope, or dust fragmentation due to collisions between bigger dust particles\footnote{The real situation
can be a lot more complicated since dust particles may be trapped at the gap edge induced by the planet and won't be flow into the circumplanetary disc.}. On the other hand, the dust replenishing rate needs to be very high to compensate the fast radial drift. For example, if the radial drift
timescale is 10$^{2}$ years while the gas accretion timescale is 10$^{5}$ years, the dust-to-gas mass ratio will be 1000 times smaller than the ISM value (0.01) even with
the continuous replenishment of dust and gas. 

4)  Or there are structures in the gaseous CPD (e.g. pressure traps) which can slow down or even trap dust particles in the disc. In the CPD simulations by \cite{Zhu2016}, 
rings and vortices have indeed been observed (e.g. Figure 3, 4 in that paper) due to the shock driven accretion. Although these features are transient, they may slow
down particle radial drift.

Another possibility for the existence of dust
is that CPDs are in the debris disc stage so that there is little gas in the disc (dust to gas mass ratio $\gtrsim$1 and dust drift is slow).
Since we are interested in CPDs that are actively accreting, this is beyond the scope of this paper. 

Overall, it is more likely to detect discs around planetary mass objects if the discs are 
accreting at a relatively high rate (surface density is higher so that the dust drift timescale is longer) and being fed continuously
(e.g.  through circumstellar discs, infalling envelope, etc.). It is less likely to find discs around isolated
planetary mass objects that are accreting at a low rate. 

On the other hand, OTS44 is isolated and does not have a high accretion rate either.  
The mm particles should have already drifted to the central planetary mass object (the right panel of Figure \ref{fig:drifttime1}). Based on the drift timescale argument,
we suggest that the ALMA detection by \cite{Bayo2017} indicates that
most dust may be at the micron-sized range, or the disk has sub-structures, or the gas surface density is very high so that the dust to gas 
mass ratio is significantly lower than 1/100. The latter scenario
can be tested if its gas mass  can be constrained from ALMA molecular line observations. 
 
Another interesting source is FW Tau b \citep{Kraus2014} which is a $\sim$10 $M_{J}$ companion 330 AU away from the central 0.2-0.3 $M_{\odot}$ binary.
ALMA has detected a strong emission ($\sim$1.78 mJy) at 1.3 mm \citep{Kraus2015}, implying 1-2 earth mass of dust in the disc.  
This amount of dust mass implies 1 M$_{J}$ mass of gas ($10\%$ of the central object mass) in the disc with the assumption that 
the dust-to-gas mass ratio is 1/100. 
Considering the gravitational instability will limit the gas mass, 
 it is unlikely that
the dust-to-gas mass ratio can be significantly lower than 1/100 in this disc.
We suggest that most of the radio emission may come from micron-sized dust in this disc, or there are structures
in the disc to slow down the particle radial drift.

\section{Conclusion}
We have constructed simple $\alpha$ disc models to study radio emission from circumplanetary discs (CPDs). 
A large disc parameter with $M_{p}\dot{M_{p}}$ from $10^{-8}M_{J}^2/yr$ to $10^{-5}M_{J}^2/yr$, $\alpha$
from $10^{-4}$ to $10^{-1}$, and $R_{out}$ from 0.1 to 1.85 AU has been explored.

We find that radio observations are more sensitive to CPDs at shorter wavelengths (e.g. ALMA Band 7 and above).
The radio emission by viscous discs is almost independent of the inner disc radius and the irradiation by the star or the planet. 
If the system is 140 pc away from us,
deep observations at ALMA Band 7 could detect CPDs around a Jupiter mass planet at 20 AU from the host star, when
the disc accretes at a rate of $\gtrsim 10^{-10} M_{\odot}/yr$ with $\alpha\lesssim 0.001$ or at a rate of $\gtrsim 10^{-9}M_{\odot}/yr$
with $\alpha\lesssim 0.01$. When the CPDs are closer or further away from the star, we can scale our calculated flux accordingly:
the flux scales as $R_{out}^{1.27}$ when the CPD is optically thick, and
as $R_{out}^{0.58}$ when it is optically thin. Radio emission from CPD Jets or winds may be detectable by ngVLA in future.  

We also discuss the radio emission from
various CPD models that were constructed to understand the Galilean satellite formation in our solar system. 
With the ``minimum mass sub-nebula'' model, ALMA has the sensitivity to detect the CPD if it is 5 AU from the central star.
If we use the ``gas-starved'' disc model, ALMA has enough sensitivity to detect the CPD if it is 20 AU away from the central star
with $\alpha<0.01$. 

Although ALMA may have enough sensitivity to detect CPDs, it is difficult to distinguish the unresolved CPDs from the background
circumstellar disc. Only if a deep gap/cavity has been carved out or the CPD is far away from the central star, we can distinguish them
from the circumstellar disc using ALMA's highest resolution. The high resolution requirement is where future long baselines array 
such as the ngVLA would provide the greatest  advantage compared to ALMA. 

Finally, we find that mm/cm dust particles drift extremely fast in CPDs. Within most of our parameter space, It only takes
$\lesssim$100-1000 years for mm dust particles to drift to the central star. 
The meter barrier in protoplanetary discs becomes the millimetre barrier in circumplanetary discs.
Thus, to detect CPDs at radio wavelengths, 
CPDs need to have a large population
of micron-sized dust, a high gas surface density, mechanisms to trap/slow down dust particles, or mechanisms to replenish mm particles efficiently. 
We have also commented on recent ALMA detection or non-detection 
of discs around planetary mass objects. Based on the particles' drift timescale argument, 
the radio emission may come from micron-sized dust only, or
the dust-to-gas mass ratio is significantly
smaller than 0.01 in these discs, or the disk has substructures to trap dust particles. These can be tested by future observations.

\section*{Acknowledgements}
Z.Z. would like to thank  Enrique Macias Quevedo, Catherine Espaillat for helpful discussions. 
The work is stimulated by the discussion of the ``Science Use Cases'' for ngVLA.
Z. Z. acknowledges support from the National Aeronautics and Space Administration 
through the Astrophysics Theory Program with Grant No. NNX17AK40G. 
A. I. acknowledges support from the National Aeronautics and Space Administration Grant
No. NNX15AB06G. 

\input{paper.bbl}
\bsp
\label{lastpage}
\end{document}